\documentclass[a4paper]{article}
\usepackage[utf8]{inputenc}
\usepackage[T1]{fontenc}
\usepackage{amsmath,amssymb,array}
\usepackage{booktabs}
\usepackage{natbib}
\usepackage{graphicx}
\usepackage{float}
\usepackage{url}
\usepackage{framed,color,verbatim}
\definecolor{shadecolor}{rgb}{.95, .95, .95}
\usepackage[a4paper,top=3cm,bottom=3cm,left=3cm,right=3cm]{geometry}
\newenvironment{example}%
   {\snugshade\verbatim}%
   {\endverbatim\endsnugshade}
\title{\texttt{stopp}: Methods for spatio-
temporal point pattern analysis, simulation, model fitting, diagnostics, and local analyses}
\author{by Nicoletta D'Angelo and Giada Adelfio}
\begin{document}



\maketitle

\section{Abstract}
The  \textbf{stopp} \texttt{R} package deals with spatio-temporal point processes which might have occurred on the Euclidean space or on some specific linear networks such as roads of a city.
The package contains functions to summarize, plot, and perform different kinds of analyses on point processes, mainly following the methods proposed in  some recent papers in the stream of scientific literature. The main topics of such works, and of the package in turn,  include
modeling, statistical inference, and simulation issues on spatio-temporal point processes on Euclidean space and linear networks, with a focus on their local characteristics. We contribute to the existing literature by collecting many of the most widespread methods for the analysis of spatio-temporal point processes into a unique package, which is intended to welcome many further proposals and extensions.

\section{Introduction}

Modelling real problems through space-time point processes is  crucial  in many scientific and engineering fields such as environmental sciences, meteorology, image analysis, seismology, astronomy, epidemiology and criminology.
The growing availability of data is a  challenging opportunity for the scientific research, aiming at  more detailed information through the application of statistical methodologies suitable for describing complex phenomena. 

The aim of the present work is to contribute to the existing literature by gathering many of the most widespread methods for the analysis of spatio-temporal point processes into a unique package, which is intended to host many further extensions.
The \textbf{stopp} \citep{R} package provides codes, related to methods and models, for analysing complex spatio-temporal point processes, proposed in the papers \cite{siino2018joint,siino2018testing,adelfio2020some,dangelo2021assessing,dangelo2021local,d2022locally}. 
The main topics include modelling, statistical inference, and simulation issues on spatial and spatio-temporal point processes, point processes on linear networks, non-Euclidean spaces.
The context of application is very broad, as the proposed methods are of interest in describing any phenomenon with a complex spatio-temporal dependence.
Some examples, include seismic events \citep{dangelo2021locall}, GPS data \citep{dangelo2021inhomogeneous},  crimes \citep{dangelo2021self}, and traffic accidents.
Moreover, local methods and models can be applied to different scientific fields and could be suitable for all those phenomena for which it makes sense to hypothesize interdependence in space and time. 



The main dependencies of the \textbf{stopp}  package are \textbf{spatstat} \cite{spatstat}, \textbf{stpp} \cite{gabriel2009second}, and  \textbf{stlnpp} \cite{moradi2020first}.
In the purely spatial context, \textbf{spatstat} is by far the most comprehensive open-source toolbox for analysing spatial point patterns, focused mainly on two-dimensional point patterns. We exploit many functions from this package when needing purely spatial tools while performing spatio-temporal analyses.
Turning to the spatio-temporal context, \textbf{stpp} represents the main reference of statistical tools for analyzing the global and local second-order properties of spatio-temporal point processes, including estimators of the space-time inhomogeneous $K$-function and pair correlation function. The package is documented in the paper \cite{gabriel:rowlingson:diggle:2013}.
While \textbf{stpp} allows for the simulation of Poisson, inhibitive and clustered patterns, the \textbf{stppSim} \citep{stppSim} package generates artificial spatio-temporal point patterns through the integration of microsimulation and agent-based models. 
Moreover, \textbf{splancs} \citep{splancs} fosters many tools for the analysis of both spatial and spatio-temporal point patterns \citep{rowlingson1993splancs,bivand2000implementing}.
Moving to spatio-temporal point patterns on linear networks, the package \textbf{stlnpp} provides tools to visualise and analyse such patterns using the first- and second-order summary statistics developed in \cite{moradi2020first,mateu2020spatio}.
Other worth-to-mention packages dealing with spatio-temporal point pattern analysis include \textbf{etasFLP} \cite{chiodi:adelfio:14}, mainly devoted to the estimation of the components of an ETAS (Epidemic Type Aftershock Sequence) model for earthquake description with the non-parametric background seismicity  estimated through FLP (Forward Likelihood Predictive) \cite{adelfio2020including}, and 
\textbf{SAPP}, 
 the Institute of Statistical Mathematics package \citep{ogata2006timsac84,ogata2006statistical}, which provides functions for the statistical analysis of series of events and seismicity.
%
Finally, we highlight some \texttt{R} packages that implement routines to simulate and fit log-Gaussian Cox processes (LGCPs). In particular, the package \textbf{stpp}  implements code to simulate spatio-temporal LGCP with a separable and non-separable covariance
structure for the Gaussian Random Field (GRF). Instead, the package \textbf{lgcp} \cite{taylor:davies:barry:15} implements code to fit LGCP models using methods
of the moments and a Bayesian inference for spatial, spatio-temporal,
multivariate and aggregated point processes. Furthermore, the minimum contrast method is used to estimate parameters assuming a separable structure of the covariance of
the GRF. Both packages do not handle for non-separable (and anisotropic)
correlation structures of the covariance structure of the GRF.

The outline of the paper is as follows.
First, we set the notation of spatio-temporal point processes, both occurring on Euclidean space and on linear networks. Then, we introduce the main functions for handling point processes objects, data, and simulations from different point process models. We then move to the Local Indicators of Spatio-Temporal Association functions, recalling their definition on the spatio-temporal Euclidean space and introducing the new functions to compute the LISTA functions on linear networks. Then, we illustrate how to perform a local test for assessing the local differences in two point patterns occurring on the same metric space.  Hence, the  functions available in the package for fitting models are illustrated, including separable Poisson process models on both the Euclidean space and networks, global and local non-separable inhomogeneous Poisson processes and LGCPs. Then, methods to perform global and local diagnostics on both models for point patterns on planar and linear network spaces are presented. The paper ends with some conclusions.

\section{Spatio-temporal point processes and their second-order properties}
\label{sec:stpp}

We consider a spatio-temporal point process with no multiple points as a random countable subset $X$ of $\mathbb{R}^2 \times \mathbb{R}$, where a point $(\textbf{u}, t) \in X$ corresponds to an event at $ \textbf{u} \in \mathbb{R}^2$ occurring at time $t \in \mathbb{R}$.
A typical realisation of a spatio-temporal point process $X$ on $\mathbb{R}^2 \times \mathbb{R}$ is a finite set $\{(\textbf{u}_i, t_i)\}^n_{
i=1}$ of distinct points within a
bounded spatio-temporal region $W \times T \subset \mathbb{R}^2 \times \mathbb{R}$, with area $\vert W\vert  > 0$ and length $\vert T\vert  > 0$, where $n \geq 0$ is not fixed in
advance. 
In this context, $N(A \times B)$ denotes the number of points of a set $(A \times B) \cap X$, where $A \subseteq W$ and $B \subseteq T$. As usual \citep{daley:vere-jones:08}, when $N(W \times T) < \infty $ with probability 1, which holds e.g. if $X$ is defined on a bounded set, we call $X$ a finite spatio-temporal point process.

For a given event $(\textbf{u}, t)$, the events that are close to $(\textbf{u}, t)$ in both space and time, for each spatial distance $r$ and time lag $h$, are given by the corresponding spatio-temporal cylindrical neighbourhood of the event $(\textbf{u}, t)$, which can be expressed by the Cartesian product as
$$
b((\textbf{u}, t), r, h) = \{(\textbf{v}, s) : \vert \vert\textbf{u} - \textbf{v}\vert \vert \leq r, \vert t - s \vert \leq h\} , \quad \quad
(\textbf{u}, t), (\textbf{v}, s) \in W \times T,
$$
where $ \vert \vert \cdot \vert \vert$ denotes the Euclidean distance in $\mathbb{R}^2$. Note that $b((\textbf{u}, t), r, h)$ is a cylinder with centre (\textbf{u}, t), radius $r$, and height $2h$.

Product densities $\lambda^{(k)}, k  \in \mathbb{N} \text{ and }  k  \geq 1 $, arguably the main tools in the statistical analysis of point processes, may be defined through the so-called Campbell Theorem (see \cite{daley:vere-jones:08}),  that constitutes an essential result in spatio-temporal point process theory, stating that, given a spatio-temporal point process $X$, for any non-negative function $f$ on $( \mathbb{R}^2 \times \mathbb{R} )^k$

\begin{equation*}
  \mathbb{E} \Bigg[ \sum_{\zeta_1,\dots,\zeta_k \in X}^{\ne} f( \zeta_1,\dots,\zeta_k)\Bigg]=\int_{\mathbb{R}^2 \times \mathbb{R}} \dots \int_{\mathbb{R}^2 \times \mathbb{R}} f(\zeta_1,\dots,\zeta_k) \lambda^{(k)} (\zeta_1,\dots,\zeta_k) \prod_{i=1}^{k}\text{d}\zeta_i,
\label{eq:campbell0}  
\end{equation*}
where $\neq$ indicates that the sum is over distinct values. In particular, for $k=1$ and $k=2$, these functions are respectively called the \textit{intensity function} $\lambda$ and the \textit{(second-order) product density} $\lambda^{(2)}$.
Broadly speaking, the intensity function describes the rate at which the events occur in the given spatio-temporal region, while the second-order product densities are used for describing spatio-temporal variability and correlations between pair of points of a pattern. They represent the point process analogues of the mean function and the covariance function of a real-valued process, respectively.
Then, the first-order intensity function is defined as 
\begin{equation*}
 \lambda(\textbf{u},t)=\lim_{\vert \text{d}\textbf{u} \times \text{d}t\vert  \rightarrow 0} \frac{\mathbb{E}[N(\text{d}\textbf{u} \times \text{d}t )]}{\vert \text{d}\textbf{u} \times \text{d}t\vert }, 
\end{equation*}
where $\text{d}\textbf{u} \times \text{d}t $ defines a small region around the point $(\textbf{u},t)$ and $\vert \text{d}\textbf{u} \times \text{d}t\vert $ is its volume. The second-order intensity function is given by
\begin{equation*}
     \lambda^{(2)}((\textbf{u},t),(\textbf{v},s))=\lim_{\vert \text{d}\textbf{u} \times \text{d}t\vert ,\vert \text{d}\textbf{v} \times \text{d}s\vert  \rightarrow 0} \frac{\mathbb{E}[N(\text{d}\textbf{u} \times \text{d}t )N(\text{d}\textbf{v} \times \text{d}s )]}{\vert \text{d}\textbf{u} \times \text{d}t\vert  \vert \text{d}\textbf{v} \times \text{d}s\vert }.
\end{equation*}
Finally, the pair correlation function
    $g((\textbf{u},t),(\textbf{v},s))=\frac{ \lambda^{(2)}((\textbf{u},t),(\textbf{v},s))}{\lambda(\textbf{u},t)\lambda(\textbf{v},s)}$
can be interpreted formally as the standardised probability density that an event occurs in each of two small volumes, $\text{d}\textbf{u} \times \text{d}t$ and $\text{d}\textbf{v} \times \text{d}s$, in the sense that for a Poisson process, $g((\textbf{u},t),(\textbf{v},s))=1.$

In this package, the focus is on second-order characteristics of spatio-temporal point patterns, with an emphasis on the $K$-function \citep{ripley:76}.
This is a measure of the distribution of the inter-point distances and captures the spatio-temporal  dependence of a point process.
%
A spatio-temporal point process is second-order intensity reweighted stationary and isotropic if its intensity function is bounded away from zero and its pair correlation function depends only on the spatio-temporal difference vector $(r,h)$, where $r= \vert \vert \textbf{u}-\textbf{v} \vert \vert $ and $h= \vert t-s \vert$ \citep{gabriel2009second}.
For a second-order intensity reweighted stationary, isotropic spatio-temporal point process, the space-time inhomogeneous $K$-function takes the form 
\begin{equation}
    K(r,h)=2 \pi \int_{-r}^{r} \int_0^{h} g(r',h')r'\text{d}r'\text{d}h'
\end{equation}
where $g(r,h)=\lambda^{(2)}(r,h)/(\lambda(\textbf{u},t)\lambda(\textbf{v},s)), r=\vert \vert\textbf{u}-\textbf{v}\vert \vert,h= \vert t-s \vert$ \citep{gabriel2009second}.
The simplest expression of an estimator of the spatio-temporal $K$-function is given  as
\begin{equation}
    \hat{K}(r,h)=\frac{1}{ \vert W  \vert  \vert   T  \vert}\sum_{i=1}^n \sum_{j > i} I(  \vert  \vert \textbf{u}_i-\textbf{u}_j  \vert  \vert \leq r, \vert t_i-t_j \vert \leq h).
    \label{eq:k}
\end{equation}
For a homogeneous Poisson process $\mathbb{E}[\hat{K}(r,h)]=\pi r^2 h$, regardless of the intensity $\lambda$.
The $K$-function  can be used as a measure of spatio-temporal clustering and interaction \citep{gabriel2009second,moller2012aspects}.
Usually, $\hat{K}(r,h)$ is compared with the theoretical $\mathbb{E}[\hat{K}(r,h)]=\pi r^2 h$. Values $\hat{K}(r,h) >  \pi r^{2} h$ suggest clustering, while $\hat{K}(r,h) < \pi r^2 h$ points to a regular pattern.


	Point processes on linear networks are recently considered to analyse events occurring on particular network structures such as the traffic accidents on a road network.
   Spatial patterns of points along a network of lines are indeed found in many applications.
The network might reflect a map of railways, rivers, electrical wires, nerve fibres, airline routes,
irrigation canals, geological faults or soil cracks \citep{baddeley2020analysing}. Observations of interest could be the locations of
traffic accidents, bicycle incidents, vehicle thefts or street crimes, and many others.
A linear network $ L=\cup_{i=1}^{n}  l_{i} \subset  \mathbb{R}^{2} $ is commonly taken as a finite union of line segments $l_i\subset  \mathbb{R}^{2}$ of positive length. 
A line segment is defined as $l_i=[u_i,v_i]=\{ku_i+(1-k)v_i: 0 \leq k \leq 1\}$, where $u_i,v_i \in \mathbb{R}^2$ are the endpoints of $l_i$. For any $i \ne j$, the intersection of $l_i$ and $l_j$ is either empty or an endpoint of both segments.\\
A spatio-temporal linear network point process is a point process on the product space $L \times T$, where $L$ is a linear network and $T$ is a subset (interval) of $\mathbb{R}$. 
	We hereafter focus on a spatio-temporal point process $X$ on a linear network $L$ with no
	overlapping points $(\textbf{u},t)$, where $\textbf{u} \in L$ is the location of an event and $t \in T (T \subseteq \mathbb{R}^+)$
	is the corresponding time occurrence of $\textbf{u}$. Note that the temporal state-space $T$ might be
	either a continuous or a discrete set. A realisation of $X$ with $n$ points is represented by
	$\textbf{x} = {(\textbf{u}_i ,t_{i} ),i = 1,\dots,n}$ where $(\textbf{u}_i ,t_{i} ) \in L \times T$.
A spatio-temporal disc with centre
	$(\textbf{u},t) \in L \times T$, network radius $r > 0$ and temporal radius $h > 0$ is defined as
	$b((\textbf{u},t ),r,h) = \{(\textbf{v},s ) : d_L (\textbf{u},\textbf{v}) \leq r , \vert t - s \vert  \leq h\}, 
(\textbf{u}, t), (\textbf{v}, s) \in L \times T $
		where  $\vert \cdot\vert $ is a numerical distance, and $d_L(\cdot,\cdot)$ stands for the appropriate distance in the network, typically taken as the shortest-path distance between any two points. The cardinality of any subset $A \subseteq L \times T, N(X \cap A) \in
	{0,1,\dots}$, is the number of points of $X$ restricted to $A$, whose expected value is denoted
	by 
	$\nu(A) = \mathbb{E}[N(X \cap A)], 
 A \subseteq L \times T,$
	where $\nu$, the intensity measure of $X$, is a locally finite product measure on $L\times T$ \citep{baddeley2006stochastic}. 
We now recall Campbell's theorem for point processes on linear networks \citep{cronie2020inhomogeneous}.
Assuming that the product densities/intensity functions $\lambda^{(k)}$ exist, for any non-negative measurable function $f(\cdot)$ on the product space $L^k$, we have
	\begin{equation}
  \mathbb{E} \Bigg[ \sum_{\zeta_1,\dots,\zeta_k \in X}^{\ne} f( \zeta_1,\dots,\zeta_k)\Bigg]=\int_{L^k}
  f(\zeta_1,\dots,\zeta_k) \lambda^{(k)} (\zeta_1,\dots,\zeta_k) \prod_{i=1}^{k}\text{d}\zeta_i.
\label{eq:campbelL}  
\end{equation}
 Assume that $X$ has an intensity function $\lambda(\cdot,\cdot)$, hence Equation \eqref{eq:campbelL} reduces to
	$\mathbb{E}[N(X \cap A)] =\int_{A} \nu(d(\textbf{u},t )) =
	\int_{A} \lambda(\textbf{u},t)d_2(\textbf{u},t), A \subseteq L \times T,$
		where $d_2 (\textbf{u},t)$ corresponds to integration over $L \times T$.
The second-order Campbell's theorem is obtained from \eqref{eq:campbelL} with $k=2$
\begin{equation}
\mathbb{E} \Bigg[ \sum_{(\textbf{u},t),(\textbf{v},s)\in X}^{\ne} f\big((\textbf{u},t),(\textbf{v},s)\big)
\Bigg] = 
\int_{L \times T} \int_{L \times T} f\big((\textbf{u},t),(\textbf{v},s)\big) \lambda^{(2)}\big((\textbf{u},t),(\textbf{v},s)\big)\text{d}_2(\textbf{u},t)\text{d}_2(\textbf{v},s).   
\label{eq:campbell}
\end{equation}
Assuming that $X$ has a second-order product density function $\lambda^{(2)} (\cdot,\cdot)$, we then obtain
	\begin{equation*}
	\mathbb{E}[N(X \cap A)N(X \cap B)] =
	\int_{A} \int_{B} 
	\lambda^{(2)} ((\textbf{u},t ),(\textbf{v},s ))d_2 (\textbf{u},t)d_2 (\textbf{v},s ), \quad A,B \subseteq L \times T.
	\end{equation*}
Finally, an important result concerns the conversion of the integration over $L \times T$  to that over $\mathbb{R} \times \mathbb{R}$ \citep{rakshit2017second}.
For any measurable function $f: L \times T \rightarrow \mathbb{R}$
\begin{equation}
    \int_{L \times T} f(\textbf{u},t)\text{d}_2(\textbf{u},t)=\int_0^{\infty} \int_0^{\infty} \sum_{\substack{ (\textbf{u},t)\in L \times T:\\ 
 d_L(\textbf{u},\textbf{v})=r,\\ 
 |t-s|=h }} f(\textbf{u},t) \text{d}r\text{d}h.
\label{eq:change}    
\end{equation}
Letting $f(\textbf{u},t) = \eta(d_L(\textbf{u},\textbf{v}), \vert t-s\vert)$
then 
$$
\int_{L \times T} \eta(d_L(\textbf{u},\textbf{v}), \vert t-s\vert) \text{d}_2(\textbf{u},t)= \int_0^{\infty} \int_0^{\infty} \eta(r,h)M((\textbf{u},t),r,h)\text{d}r \text{d}h
$$
where $M((\textbf{u},t),r,h)$ is the number of points lying exactly at the shortest-path distance $r \geq 0$ and the time distance $h \geq 0$ away from $(\textbf{u},t)$.

\section{Main functions for handling point processes objects, data, and simulations}\label{sec:main}

The \texttt{stp} function creates a \texttt{stp} object as a dataframe with three columns: \texttt{x}, \texttt{y}, and \texttt{t}.  If  the linear network \texttt{L}, of class \texttt{linnet}, is also provided, a \texttt{stlp} object is created instead.
The methods for this class of objects: (1) print the main information on the spatio-temporal point pattern stored in the \texttt{stp} object: the number of points, the enclosing spatial window, the temporal time period; (2) print the summary statistics of the spatial and temporal coordinates of the spatio-temporal point pattern stored in the \texttt{stp} object; (3) plot the point pattern stored in the \texttt{stp} object given in input, in a three-panel plot representing the 3Dplot of the coordinates, and the marginal spatial and temporal coordinates.
\begin{example}
> set.seed(12345)
> rpp1 <- stpp::rpp(lambda = 200, replace = FALSE)
> is.stp(rpp1)
[1] FALSE

> stp1 <- stp(cbind(rpp1$xyt[, 1], rpp1$xyt[, 2], rpp1$xyt[, 3]))
> is.stp(stp1)
[1] TRUE

> stp1
Spatio-temporal point pattern 
208 points 
Enclosing window: rectangle = [0.0011366, 0.9933775] x [0.0155277, 0.9960438] units
Time period: [0.004, 0.997] 
\end{example}
%
Some functions are implemented to convert the \texttt{stp} and \texttt{stlp} classes to those of the \textbf{stpp} and \textbf{stlnpp} packages, and vice-versa.


Moreover, the package is furnished with the \texttt{greececatalog} dataset in the \texttt{stp} format containing the catalog of 
Greek earthquakes of magnitude at least 4.0 from year 2005 to year 2014,
analysed by mean of local log-Gaussian Cox processes in \cite{dangelo2021locall}
and \cite{d2022locally}.
Data come from the Hellenic Unified Seismic Network (H.U.S.N.).
The same data have been analysed in \cite{siino2017spatial} by hybrids of Gibbs models,
and more recently by \cite{gabriel2022mapping}.
\begin{example}
> plot(greececatalog, tcum = TRUE)
\end{example}
\begin{figure}[H]
	\centering
	\includegraphics[width=.8\textwidth]{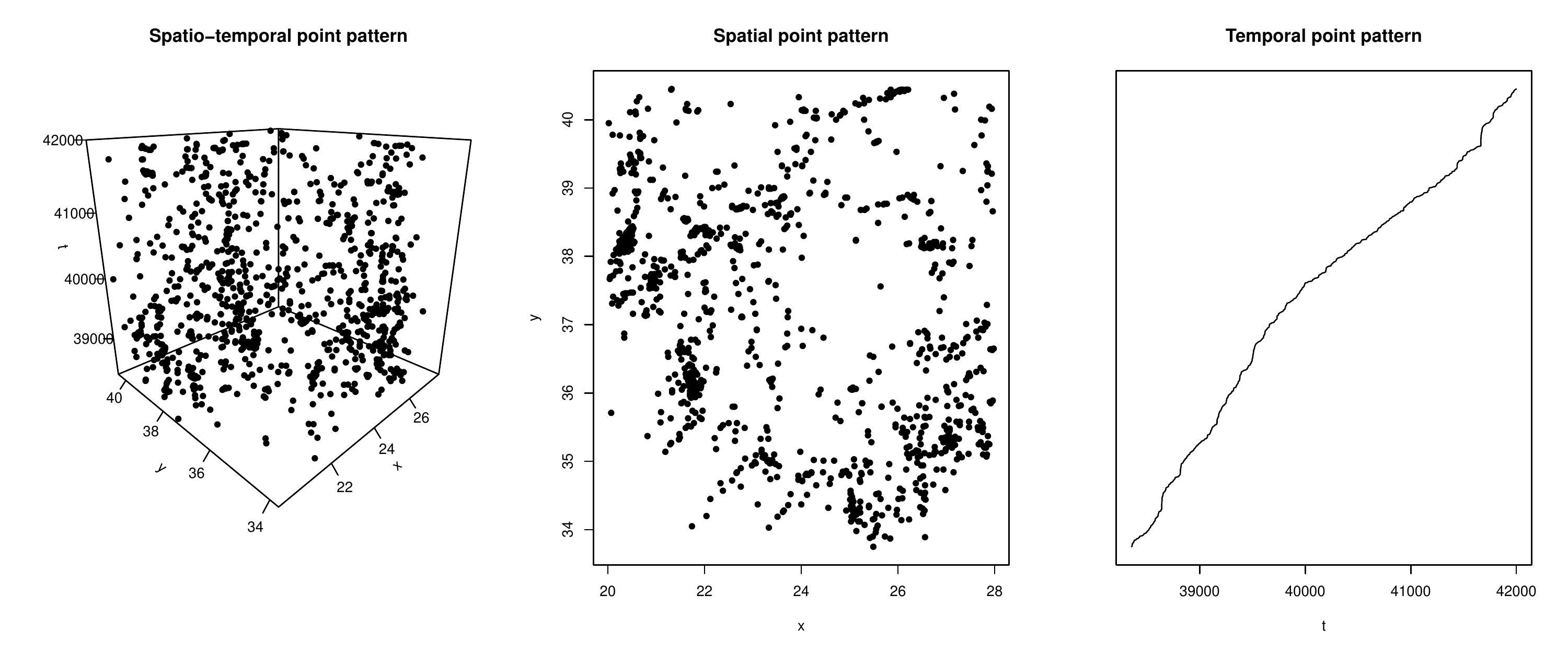}
	\caption{Plots of Greek data.}
	\label{fig:p2}
\end{figure}
A dataset of crimes occurred in Valencia, Spain, in 2019 is available, together with the linear
network of class \texttt{linnet} of the Valencian roads, named  \texttt{valenciacrimes} and \texttt{valencianet}, respectively.

Finally, the linear network of class \texttt{linnet} of the roads of Chicago (Illinois, USA) close to the University of Chicago, is also available.
It represents the linear network of the Chicago dataset published and analysed in \cite{ang2012geometrically}. The network adjacency
matrix is stored as a sparse matrix.

Moving to simulations, the \texttt{rstpp} function creates a \texttt{stp} object, simulating a spatio-temporal Poisson point pattern, following either a homogeneous or inhomogeneous intensity.
\begin{example}
> h1 <- rstpp(lambda = 500, nsim = 1, seed = 2, verbose = TRUE)
> plot(h1, tcum = TRUE)
\end{example}
\begin{figure}[H]
	\centering
	\includegraphics[width=.8\textwidth]{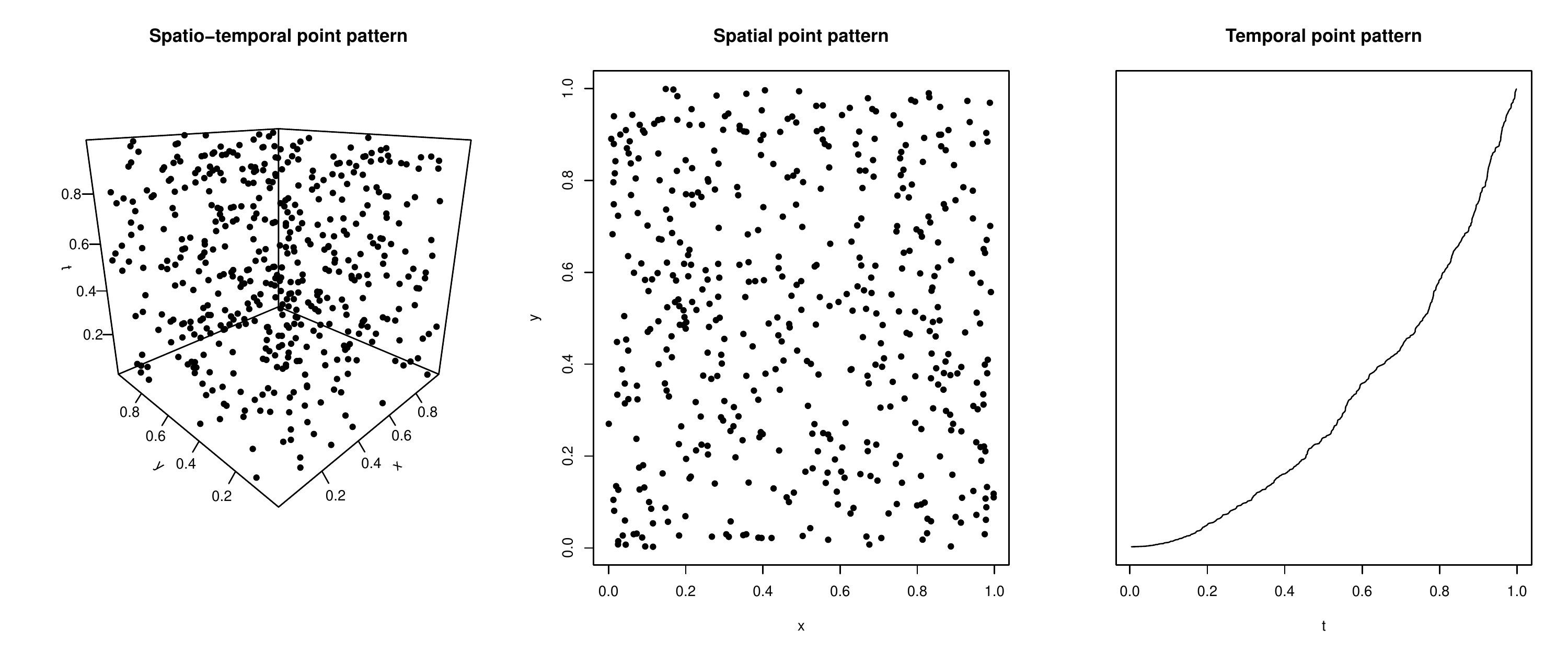}
	\caption{Simulated homogeneous point pattern.}
	\label{fig:p3}
\end{figure}
\begin{example}
> inh <- rstpp(lambda = function(x, y, t, a) {exp(a[1] + a[2]*x)}, par = c(2, 6),
             nsim = 1, seed = 2, verbose = TRUE)
> plot(inh, tcum = TRUE)
\end{example}
\begin{figure}[H]
	\centering
	\includegraphics[width=.8\textwidth]{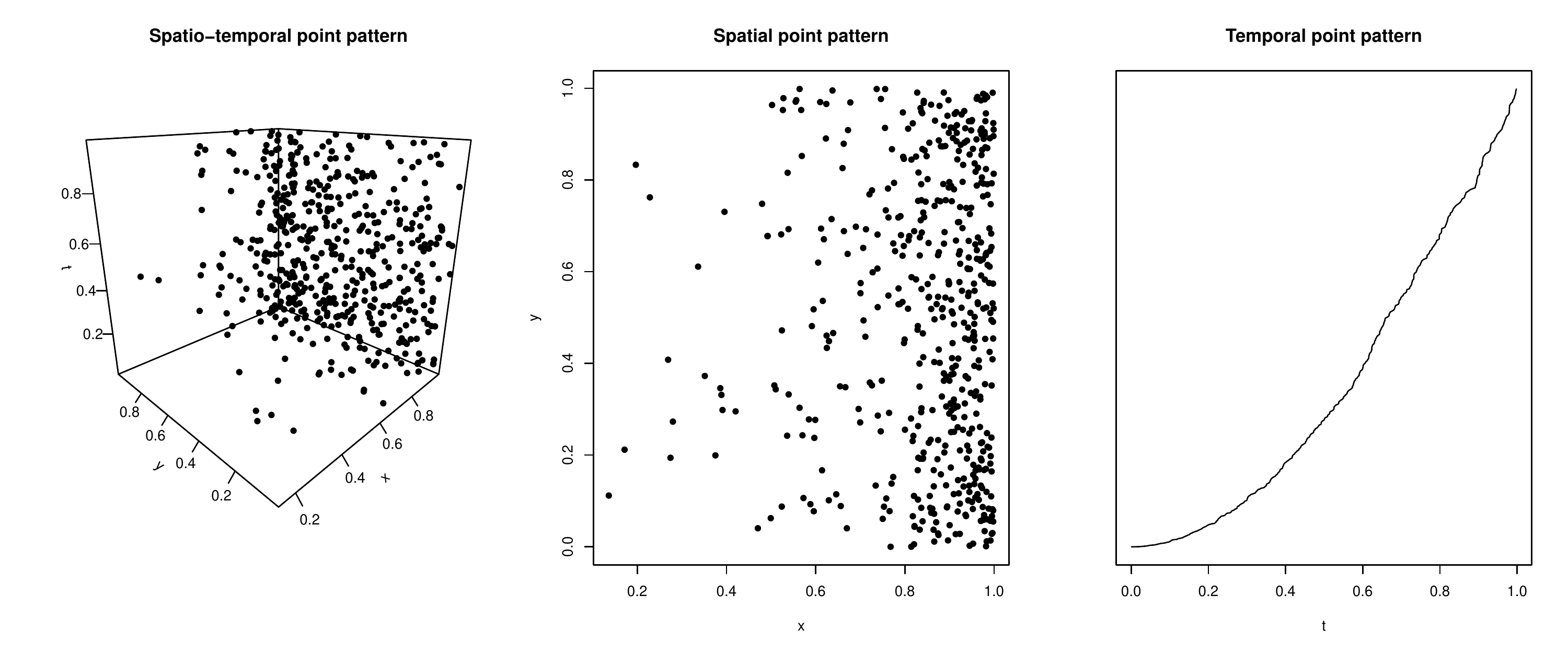}
	\caption{Simulated inhomogeneous point pattern.}
	\label{fig:p4}
\end{figure}
The \texttt{rstlpp} function creates a \texttt{stlp} object instead, simulating a spatio-temporal Poisson point pattern 
on a linear network.
%
%
%
%
%
Furthermore, the \texttt{rETASp} function creates a \texttt{stp} object, simulating a spatio-temporal ETAS (Epidemic Type Aftershock Sequence) process.
%
%
It follows the generating scheme for simulating a pattern from an ETAS process \citep{ogata:1988likelihood} with conditional intensity function (CIF) as in \cite{adelfio2020including}.
The \texttt{rETAStlp} function creates a \texttt{stlp} object, simulating a spatio-temporal ETAS  process on a linear network. The simulation scheme previously introduced is adapted for the space location of events to be constrained on a linear network, being firstly introduced and employed for simulation studies in \cite{dangelo2021assessing}.

\section{Local Indicators of Spatio-Temporal Association functions}
\label{sec:lista}

Local Indicators of Spatio-Temporal Association (LISTA) are a set of functions that are individually associated with each one of the points of the point pattern, and can provide information about the local behaviour of the pattern.
 This operational definition of local indicators was introduced by \cite{anselin:95} for the spatial case, and extended by \cite{siino2018testing} to the spatio-temporal context.\\
 If $\lambda^{(2)i}(\cdot,\cdot)$ denotes the local version of the spatio-temporal product density for the event $(\textbf{u}_i,t_i)$, 
then, for fixed $r$ and $h$, it holds that
\begin{equation}
    \hat{\lambda}^{(2)}_{\epsilon,\delta}(r,h)=\frac{1}{n-1}\sum_{i=1}^n\hat{\lambda}^{(2)i}_{\epsilon,\delta}(r,h),
    \label{eq:op}
\end{equation}
where $
    \hat{\lambda}^{(2)i}_{\epsilon,\delta}(r,h)=\frac{n-1}{4\pi r  \vert W  \times  T \vert}\sum_{j\ne i}\kappa_{\epsilon,\delta}(  \vert  \vert \textbf{u}_i-\textbf{v}_j  \vert  \vert -r, \vert t_i-s_j \vert -h),
$
with $r>\epsilon>0$ and $h>\delta>0$, and $\kappa$ a kernel function with  spatial and  temporal bandwidths $\epsilon$ and $\delta$, respectively.
Any second-order spatio-temporal summary statistic that satisfies the operational definition in \eqref{eq:op}, which means that the sum of spatio-temporal local indicator functions is proportional to the global statistic, can be called a LISTA statistic \citep{siino2018testing}.\\
In \cite{adelfio2020some}, local versions of both the homogeneous and inhomogeneous spatio-temporal $K$-functions on the Euclidean space  are introduced.
Defining an estimator of the overall intensity by $\hat{\lambda}=n/(\vert W \vert \vert T \vert)$, they propose the local version of \eqref{eq:k}  for the i-th event $(\textbf{u}_i,t_i)$ 
\begin{equation}
    \hat{K}^i(r,h)=\frac{1}{\hat{\lambda}^2  \vert W  \vert  \vert T \vert}\sum_{(\textbf{u}_i,t_i)\ne (\textbf{v},s)} I(  \vert  \vert \textbf{u}_i-\textbf{v}  \vert  \vert\leq r,\vert t_i-s\vert \leq h)
    \label{eq:kl}
\end{equation}
and the inhomogeneous version 
\begin{equation}
    \hat{K}^i_{I}(r,h)=\frac{1}{  \vert W  \vert  \vert T  \vert}\sum_{(\textbf{u}_i,t_i)\ne (\textbf{v},s)} \frac{I(||\textbf{u}_i-\textbf{v} \vert  \vert \leq r,\vert t_i-s\vert  \leq h)}{\hat{\lambda}(\textbf{u}_i,t_i)\hat{\lambda}(\textbf{v},s)},
    \label{eq:kinhl}
\end{equation}
with $(\textbf{v},s)$ being the spatial and temporal coordinates of any other point.
  The authors extended the spatial weighting approach of \cite{veen2006assessing} to spatio-temporal local second-order statistics, proving that the inhomogeneous second-order statistics behave as the corresponding homogeneous ones, basically proving that the expectation of both \eqref{eq:kl} and \eqref{eq:kinhl} is equal to $\pi r^2 h$.

\subsection{LISTA on linear networks}

The functions \texttt{localSTLKinhom} and \texttt{localSTLginhom} implement the inhomogeneous LISTA functions proposed in \cite{dangelo2021local}.
The \textit{local spatio-temporal inhomogeneous}
   K-function for the i-th event $(\boldsymbol{u}_i,t_i)$ on a linear network
 is $$\hat{K}^i_{L,I}(r,h)=\frac{1}{ \vert L  \vert  \vert T  \vert}\sum_{(\boldsymbol{u}_i,t_i)\ne (\boldsymbol{v},s)} \frac{I\{ d_L(\boldsymbol{u}_i,\boldsymbol{v})<r,\vert t_i-s\vert <h\} }{\hat{\lambda}(\boldsymbol{u}_i,t_i)\hat{\lambda}(\boldsymbol{v},s)M((\boldsymbol{u}_i,t_i),d_L(\boldsymbol{u}_i,\boldsymbol{v}),\vert t_i-s\vert )},$$
  and the corresponding \textit{local pair correlation function} (pcf)
   $$\hat{g}^i_{L,I}(r,h)=\frac{1}{ \vert L  \vert  \vert T  \vert}\sum_{(\boldsymbol{u}_i,t_i)\ne (\boldsymbol{v},s)} \frac{\kappa( d_L(\boldsymbol{u}_i,\boldsymbol{v})-r)\kappa(\vert t_i-s\vert -h) }{\hat{\lambda}(\boldsymbol{u}_i,t_i)\hat{\lambda}(\boldsymbol{v},s)M((\boldsymbol{u}_i,t_i),d_L(\boldsymbol{u}_i,\boldsymbol{v}),\vert t_i-s\vert )},$$
 with
$$D(X) = \frac{n-1}{ \vert L  \vert  \vert T  \vert}\sum_{i=1}^n\sum_{i \ne j}\frac{1}{\hat{\lambda}(\textbf{u}_i,t_i)\hat{\lambda}(\textbf{u}_j,t_j)}$$
 normalization factor. This leads to the unbiased estimators $\frac{1}{D(X)}\hat{K}^i_{L,I}(r,h)$ and
$\frac{1}{D(X)}\hat{g}^i_{L,I}(r,h)$. 
 
 The homogeneous versions \citep{dangelo2021assessing} can be obtained by weighting the second-order
 summary statistics (either K or pcf) by a constant intensity
$\hat{\lambda}=n/( \vert L \vert \vert T \vert)$, giving 
  $$\hat{K}_L^i(r,h)=\frac{1}{\hat{\lambda}^{2} \vert L  \vert  \vert T  \vert}\sum_{(\boldsymbol{u}_i,t_i)\ne (\boldsymbol{v},s)} \frac{I\{ d_L(\boldsymbol{u}_i,\boldsymbol{v})<r,\vert t_i-s\vert <h\} }{M((\boldsymbol{u}_i,t_i),d_L(\boldsymbol{u}_i,\boldsymbol{v}),\vert t_i-s\vert )},$$
 and
$$\hat{g}_L^i(r,h)=\frac{1}{\hat{\lambda}^{2} \vert L  \vert  \vert T  \vert}\sum_{(\boldsymbol{u}_i,t_i)\ne (\boldsymbol{v},s)} \frac{\kappa( d_L(\boldsymbol{u}_i,\boldsymbol{v})-r)\kappa(\vert t_i-s\vert -h) }{M((\boldsymbol{u}_i,t_i),d_L(\boldsymbol{u}_i,\boldsymbol{v}),\vert t_i-s\vert )}.$$
These can be computed easily with the functions  \texttt{localSTLKinhom} and \texttt{localSTLKginhom}, by imputing a lambda vector of constant intensity values, the same for each point.

The proposed functions are the local counterparts of \texttt{STLKinhom} and \texttt{STLKginhom} by \cite{moradi2020first}, available in the \texttt{stlnpp} package \citep{stlnpp}. 

\begin{example}
> set.seed(10)
> X <- stlnpp::rpoistlpp(.2, a = 0, b = 5, L = stlnpp::easynet)
> lambda <- density(X, at = "points")
> x <- as.stlp(X)
> k <- localSTLKinhom(x, lambda = lambda, normalize = TRUE)

## select an individual point
> j = 1
> k[[j]]

## plot the lista function and compare it with its theoretical value
> inhom <- list(x = k[[j]]$r, y = k[[j]]$t, z = k[[j]]$Kinhom)
> theo <- list(x = k[[j]]$r, y = k[[j]]$t, z = k[[j]]$Ktheo)
> diff <- list(x = k[[j]]$r, y = k[[j]]$t, z = k[[j]]$Kinhom - k[[j]]$Ktheo)
> oldpar <- par(no.readonly = TRUE)
> par(mfrow = c(1, 3))
> fields::image.plot(inhom, main= "Kinhom", col = hcl.colors(12, "YlOrRd", rev = FALSE), 
                   xlab = "Spatial distance", ylab = "Temporal distance")
> fields::image.plot(theo, main = "Ktheo", col = hcl.colors(12, "YlOrRd", rev = FALSE), 
                   xlab = "Spatial distance", ylab = "Temporal distance")
> fields::image.plot(diff, main = "Kinhom - Ktheo", col = hcl.colors(12, "YlOrRd", rev = FALSE), 
                   xlab = "Spatial distance", ylab = "Temporal distance")
> par(oldpar)
\end{example}
\begin{figure}[H]
	\centering
	\includegraphics[width=.9\textwidth]{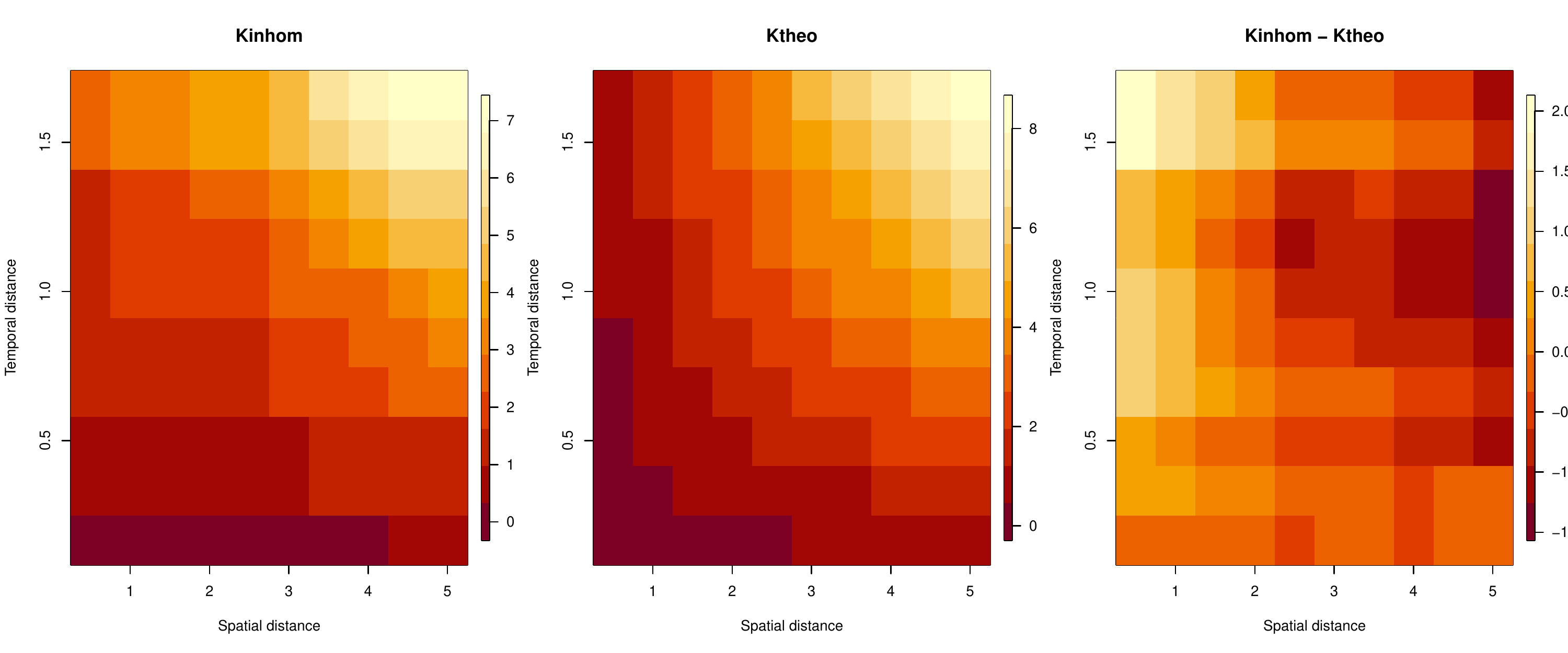}
	\caption{Observed vs theoretical K-function.}
	\label{fig:p5}
\end{figure}

\subsection{Local test for assessing the second-order differences between of two point patterns}\label{sec:test}

The function \texttt{localtest} performs the permutation test of the local structure of spatio-temporal point pattern data, proposed in \cite{siino2018testing}.
The network counterpart is also implemented, following \cite{dangelo2021assessing}.
This test detects local differences in the second-order structure of two observed point patterns $\textbf{x}$ and $\textbf{z}$  
 occurring on the same space-time region.
 This procedure was firstly introduced in \cite{moraga:montes:11} for the purely spatial  case, and then extended in 
 the spatio-temporal context by \cite{siino2018testing}. Finally,
 test has been made suitable also for spatio-temporal point patterns
 with spatial domain coinciding with a linear network by \cite{dangelo2021assessing}.
 In general, for each point $(\textbf{u},t)$ in the spatio-temporal observed
 point pattern $\textbf{x}$, we test
 $$
 \begin{cases}
 \mathcal{H}_{0}: & \text{no difference in the second-order local  structure of }  (\textbf{u},t) \quad   \text{   w.r.t  } \quad \{  \{  \textbf{x} \setminus    (\textbf{u},t)   \} \cup  \textbf{z} \}\\
 \mathcal{H}_{1}: &  \text{significant difference in the second-order local  }   \text{structure of} (\textbf{u},t) \quad \text{   w.r.t  } \quad \{  \{  \textbf{x} \setminus    (\textbf{u},t)   \} \cup  \textbf{z} \}
 \end{cases}$$
 
 The sketch of the test is as follows:
\begin{enumerate}
    \item Set $k$ as the number of permutations
        \item For each point  $(\textbf{u}_i,t_i) \in \textbf{x}, i = 1, \ldots, n$:
        \begin{itemize}
            \item Estimate the LISTA function  $\hat{L}^{(i)}(r,h)$ 
            \item Compute the local  deviation test  
   $$T^i=\int_{0}^{t_0} \int_{0}^{r_0} \Big(
     \hat{L}^{(i)}(r,h)- \hat{L}^{-(i)}_{H_0}(r,h)
     \Big)^2 \text{d}r \text{d}h,$$
     where  $\hat{L}^{-(i)}_{H_0}(r,h)$
      is the LISTA function for the $i^{th}$ point,
       averaged over the $j=1,\dots,k$ permutations 
            \item Compute a $p$-value as 
       $p^i=\sum_{j=1}^{k}  \textbf{1}(T^{i,j}_{H_0} \geq T^i)/k$ 
        \end{itemize}
\end{enumerate}

    The test ends providing a vector $p$ of  $p$- values, one for each point
    in $\textbf{x}$. 
    
  If the test is  performed for spatio-temporal point patterns as in
    \cite{siino2018testing}, that is, on an object of class \texttt{stp}, the LISTA
    functions $\hat{L}^{(i)}$ employed are the local $K$-functions of 
    \cite{adelfio2020some}, computed by the function 
  \texttt{KLISTAhat} 
  of the \textbf{stpp} package \citep{gabriel:rowlingson:diggle:2013}
  .
  If the function is applied to a \texttt{stlp} object, that is, on two spatio-temporal
  point patterns observed on the same linear network \texttt{L}, 
  the local $K$-functions
 used are the ones proposed in \cite{dangelo2021assessing}, documented
 in  \texttt{localSTLKinhom}. 
Details on the performance of the test are found in \cite{siino2018testing} and
\cite{dangelo2021assessing} for Euclidean and network spaces, respectively.
 Alternative LISTA functions that can be employed to run the test are  \texttt{LISTAhat} of \textbf{stpp} and \texttt{localSTLginhom} of \textbf{stopp}, that is, the pcfs on Euclidean space and 
linear networks respectively.

The methods for this class of objects: (1) print the main information on the result of the local permutation test performed with \texttt{localtest} on either a \texttt{stp} or \texttt{stlp} object: whether the local test was run on point patterns lying on a linear network or not; the number of points in the background \texttt{X} and alternative \texttt{Z} patterns; the number of points in \texttt{X} which exhibit local differences in the second-order structure with respect to \texttt{Z}, according to the performed test;
(2) plot the result of the local permutation test performed with \texttt{localtest}: it highlights the points of the background pattern \texttt{X}, which exhibit local differences in the second-order structure with respect to \texttt{Z}, according to the previously performed test. The remaining points of \texttt{X} are also represented; it also shows the underlying linear network, if the local test has been applied to point patterns occurring on the same linear network, that is, if \texttt{localtest} has been applied to a \texttt{stlp} object. In the following, we provide an example of two point processes, both occurring on the unit cube.

\begin{example}
## background pattern
> set.seed(12345)
> X <- rstpp(lambda = function(x, y, t, a) {exp(a[1] + a[2]*x)}, par = c(.05, 4),
           nsim = 1, seed = 2, verbose = TRUE)

## alternative pattern
> set.seed(12345)
> Z <- rstpp(lambda = 25, nsim = 1, seed = 2, verbose = TRUE)

## run the local test
> test <- localtest(X, Z, method = "K", k = 9, verbose = FALSE)

> test

Test for local differences between two 
spatio-temporal point patterns 
--------------------------------------
Background pattern X: 17  
Alternative pattern Z: 20  
  
1 significant points at alpha = 0.05
> plot(test)
\end{example}
\begin{figure}[H]
	\centering
	\includegraphics[width=\textwidth]{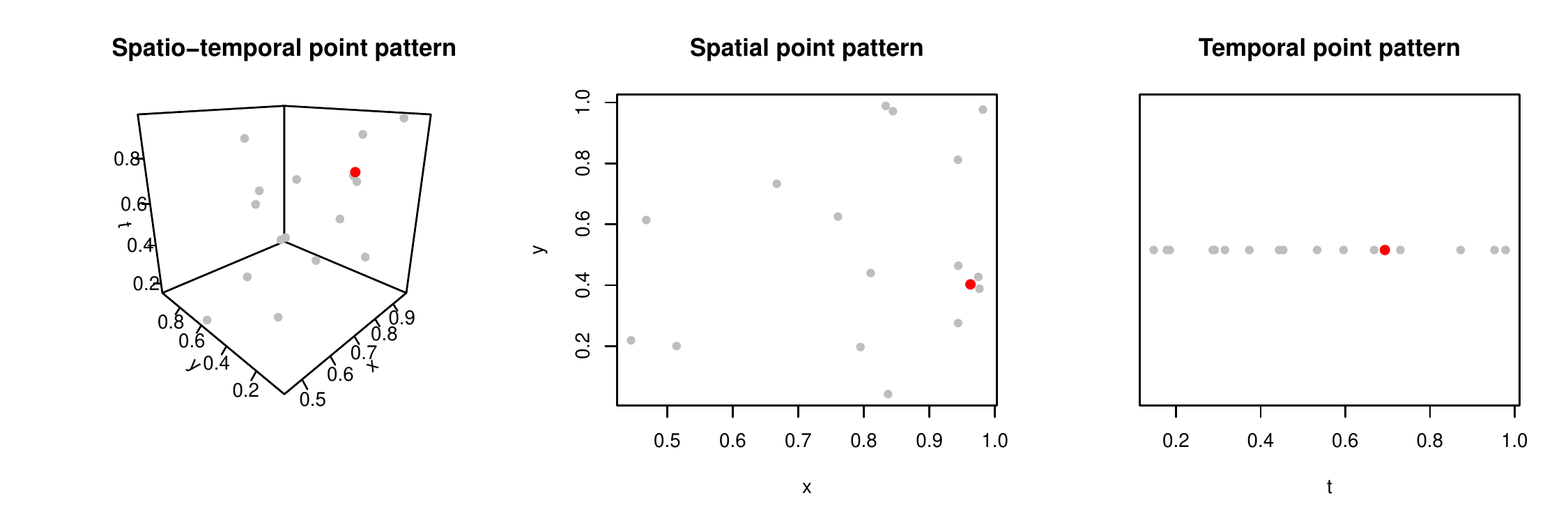}
	\caption{Output of the local test.}
	\label{fig:p14}
\end{figure}
\section{Model fitting}\label{sec:models}

	The description of the observed point pattern intensity is a crucial issue dealing with spatio-temporal point pattern data, and specifying a statistical model is a very effective way compared to analyzing data by calculating summary statistics. Formulating and adapting a statistical model to the data allows taking into account effects that otherwise could introduce distortion in the analysis \citep{baddeley2015spatial}. In this section, we outline the main functions to fit different specifications of inhomogeneous spatio-temporal Poisson process models.

\subsubsection{Spatio-temporal Poisson point processes with separable intensity}


	When dealing with intensity estimation for spatio-temporal point processes, it is quite common to assume that the intensity function $\lambda(\textbf{u},t)$ is separable \citep{diggle2013statistical,gabriel2009second}. Under this assumption,   the intensity function is given by the product 
 \begin{equation}
   \lambda(\textbf{u},t)={\lambda}(\textbf{u}){\lambda}(t)
   \label{eq:sep}
 \end{equation}
	where ${\lambda}(\textbf{u})$ and ${\lambda}(t)$ are non-negative functions on $W$ and $T$, respectively  \citep{gabriel2009second}.
 Under this assumption, any non-separable effects are interpreted as second-order, rather than first-order. Suitable  estimates  of $\lambda(\textbf{u})$ and $\lambda(t)$ in  \eqref{eq:sep}    depend  on  the  characteristics  of each application. The functions here implemented use a combination of a parametric spatial point pattern model, potentially depending on the spatial coordinates and/or spatial covariates, and a parametric log-linear model for the temporal component. Also, non-parametric kernel estimate form(s) are legit but still not implemented.
 The spatio-temporal intensity is therefore obtained by multiplying the purely spatial and purely temporal intensities, previously  fitted separately. The resulting intensity is normalised, to make the estimator unbiased, making the expected number of points
		$$\mathbb{E}\bigg[ \int_{W \times T}  \hat{\lambda}(\textbf{u},t)d_2(\textbf{u},t) \bigg] = \int_{W \times T} \lambda(\textbf{u},t)d_2(\textbf{u},t)=n,$$

and the final intensity function is obtained as
		$$\hat{\lambda}(\textbf{u},t)=\frac{\hat{\lambda}(\textbf{u})\hat{\lambda}(t)}{\int_{W \times T} \hat{\lambda}(\textbf{u},t)d_2(\textbf{u},t)}.$$


The function \texttt{sepstppm} fits such a separable spatio-temporal Poisson process model.
The function \texttt{plot.sepstppm} shows the fitted intensity, displayed both in space and in space and time.

\begin{example}
> df1 <- valenciacrimes[valenciacrimes$x < 210000 & valenciacrimes$x > 206000
    & valenciacrimes$y < 4377000 & valenciacrimes$y > 4373000, ]

> mod1 <- sepstppm(df1, spaceformula = ~x * y, timeformula = ~ crime_hour + week_day)
\end{example}

 For linear network point patterns,  non-parametric estimators of the intensity function $\lambda(\cdot,\cdot)$ have been proposed \citep{mateu2020spatio}, suggesting any variation of the distribution of the process over its state-space $L \times T$.
	A kernel-based intensity estimator for spatio-temporal linear network point processes, based on the first-order separability assumption, considered in \cite{moradi2020first}, is obtainable with the package \textbf{stnlpp}.
The functions \texttt{sepstlppm} and \texttt{plot.sepstlppm} implement the network counterparts of the spatio-temporal Poisson point process with separable intensity and fully parametric specification.
\begin{example}
> mod1 <- sepstlppm(valenciacrimes[1:2500, ], spaceformula = ~x,
    timeformula = ~ crime_hour + week_day, L = valencianet) 
\end{example}
\subsubsection{Global inhomogeneous spatio-temporal Poisson processes trough quadrature scheme}

For a non-separable spatio-temporal specification, we assume that the template model is a Poisson process, with a parametric intensity or rate function 
\begin{equation}
    \lambda(\textbf{u}, t; \theta), \quad  \textbf{u} \in
W,\quad  t \in T, \quad \theta \in \Theta.
\label{eq:pois}
\end{equation}
The log-likelihood of the template model is 
 $$\log L(\theta) = \sum_i
 \lambda(\textbf{u}_i, t_i; \theta) - \int_W\int_T
 \lambda(\textbf{u}, t; \theta) \text{d}t\text{d}u$$
 up to an additive constant, where the sum is over all points $\textbf{u}_i$
  in the spatio-temporal point process $X$.
  We might consider intensity models of log-linear form
\begin{equation}
   \lambda(\textbf{u}, t; \theta) = \exp(\theta Z(\textbf{u}, t) + B(\textbf{u},t )), \quad
\textbf{u} \in W,\quad  t \in T
\label{eq:glo_mod}
\end{equation}
where $Z(\textbf{u}, t)$ is a vector-valued covariate function, and $B(\textbf{u}, t)$ is a scalar offset. 
In  point process theory, the  variables $Z(\textbf{u}, t)$ are referred to as spatio-temporal covariates. Their observable values are assumed to be knowable, at least in principle, at each location in the spatio-temporal window. 
For inferential purposes, their values must be known at each point of the data point pattern and at least at some other locations. 
This is the reason why we first implmented the dependence of the intensity function  $\lambda(\textbf{u}, t; \theta)$ on the space and time coordinates first.\\
The \texttt{stppm} function fits a Poisson process model to an observed spatio-temporal point pattern stored in a \texttt{stp} object,  assuming the template model \eqref{eq:pois}.

Estimation is performed by fitting a \texttt{glm} using a spatio-temporal version of the quadrature scheme by \cite{berman1992approximating}.
We use a finite quadrature approximation
 to the log-likelihood. Renaming the data points as $\textbf{x}_1,\dots , 
 \textbf{x}_n$ with $(\textbf{u}_i,t_i) = \textbf{x}_i$ for $i = 1, \dots , n$,
 then generate $m$  additional 'dummy points' $(\textbf{u}_{n+1},t_{n+1})
 \dots , (\textbf{u}_{m+n},t_{m+n})$ to
 form a set of $n + m$ quadrature points (where $m > n$). 
 Then we determine quadrature weights $a_1, \dots , a_m$
 so that a Riemann sum can approximate integrals in the log-likelihood
 $$ \int_W \int_T \lambda(\textbf{u},t;\theta)\text{d}t\text{d}u \approx \sum_{k = 1}^{n + m}a_k\lambda(\textbf{u}_{k},t_{k};\theta)$$
 where $a_k$ are the quadrature weights such that
 $\sum_{k = 1}^{n + m}a_k = l(W \times T)$ where $l$ is the Lebesgue measure.
 Then the log-likelihood  of the template model can be approximated by
 $$ \log L(\theta)   \approx \sum_i \log \lambda(\textbf{x}_i; \theta) +\sum_j(1 - \lambda(\textbf{u}_j,t_j; \theta))a_j=\sum_je_j \log \lambda(\textbf{u}_j, t_j; \theta) + (1 - \lambda(\textbf{u}_j, t_j; \theta))a_j$$
 where $e_j = 1\{j \leq n\}$ is the indicator that equals $1$ if 
 $u_j$ is a data point. Writing $y_j = e_j/a_j$ this becomes
 $$ \log L(\theta) \approx
 \sum_j
 a_j
 (y_j \log \lambda(\textbf{u}_j, t_j; \theta) - \lambda(\textbf{u}_j, t_j; \theta))
 +
   \sum_j
 a_j.$$
 Apart from the constant $\sum_j a_j$, this expression is formally equivalent
  to the weighted log-likelihood of
 a Poisson regression model with responses $y_j$ and means 
 $\lambda(\textbf{u}_j,t_j; \theta) = \exp(\theta Z(\textbf{u}_j,t_j) +
  B(\textbf{u}_j,t_j))$.
   This is
 maximised by this function by using standard GLM software. 
 In detail, we define the spatio-temporal quadrature scheme by considering a 
 spatio-temporal 
 partition of $W \times T$ into cubes $C_k$ of equal volume $\nu$,
  assigning the weight $a_k=\nu/n_k$ 
  to each quadrature point (dummy or data) where $n_k$ is the number of 
  points that lie in the same cube as the point $u_k$ \citep{raeisi2021spatio}. 
 The number of dummy points should be sufficient for an accurate estimate of the 
 likelihood. Following \cite{baddeley2000non} and \cite{raeisi2021spatio}, 
 we start with a number of dummy points $m \approx 4 n$, increasing it until 
 $\sum_k a_k = l(W \times T)$.

The \texttt{AIC.stppm} and \texttt{BIC.stppm} functions return the $AIC = 2k - 2 \log(\hat{L})$ and $BIC = k\log{n} - 2 \log(\hat{L})$ of a point process 
model fitted through the 
function \texttt{stppm} applied to an observed
spatio-temporal point pattern of class \texttt{stp}. 
As the model returned by \texttt{stppm} is fitted through a quadrature scheme, 
the log-likelihood is computed through the quantity
$$- \log{L(\hat{\theta}; \boldsymbol{x})} = \frac{D}{2} + \sum_{j = 1}^{n}I_j\log{w_j}+n(\boldsymbol{x}).$$

\begin{example}
## Homogeneous
> set.seed(2)
> ph <- rstpp(lambda = 200, nsim = 1, seed = 2, verbose = TRUE)
> hom1 <- stppm(ph, formula = ~ 1)

> hom1
Homogeneous Poisson process 
with Intensity: 202.093

Estimated coefficients: 
(Intercept) 
      5.309 

## plot(hom1) won't show any plot, due to the constant intensity

> coef(hom1)
(Intercept) 
   5.308728 
   
## Inhomogeneous
> set.seed(2)
> pin <- rstpp(lambda = function(x, y, t, a) {exp(a[1] + a[2]*x)}, par = c(2, 6),
            nsim = 1, seed = 2, verbose = TRUE)
1.
> inh1 <- stppm(pin, formula = ~ x)

> inh1
Inhomogeneous Poisson process 
with Trend: ~x

Estimated coefficients: 
(Intercept)           x 
      2.180       5.783 
      
> plot(inh1)
\end{example}
\begin{figure}[H]
	\centering
	\includegraphics[width=.8\textwidth]{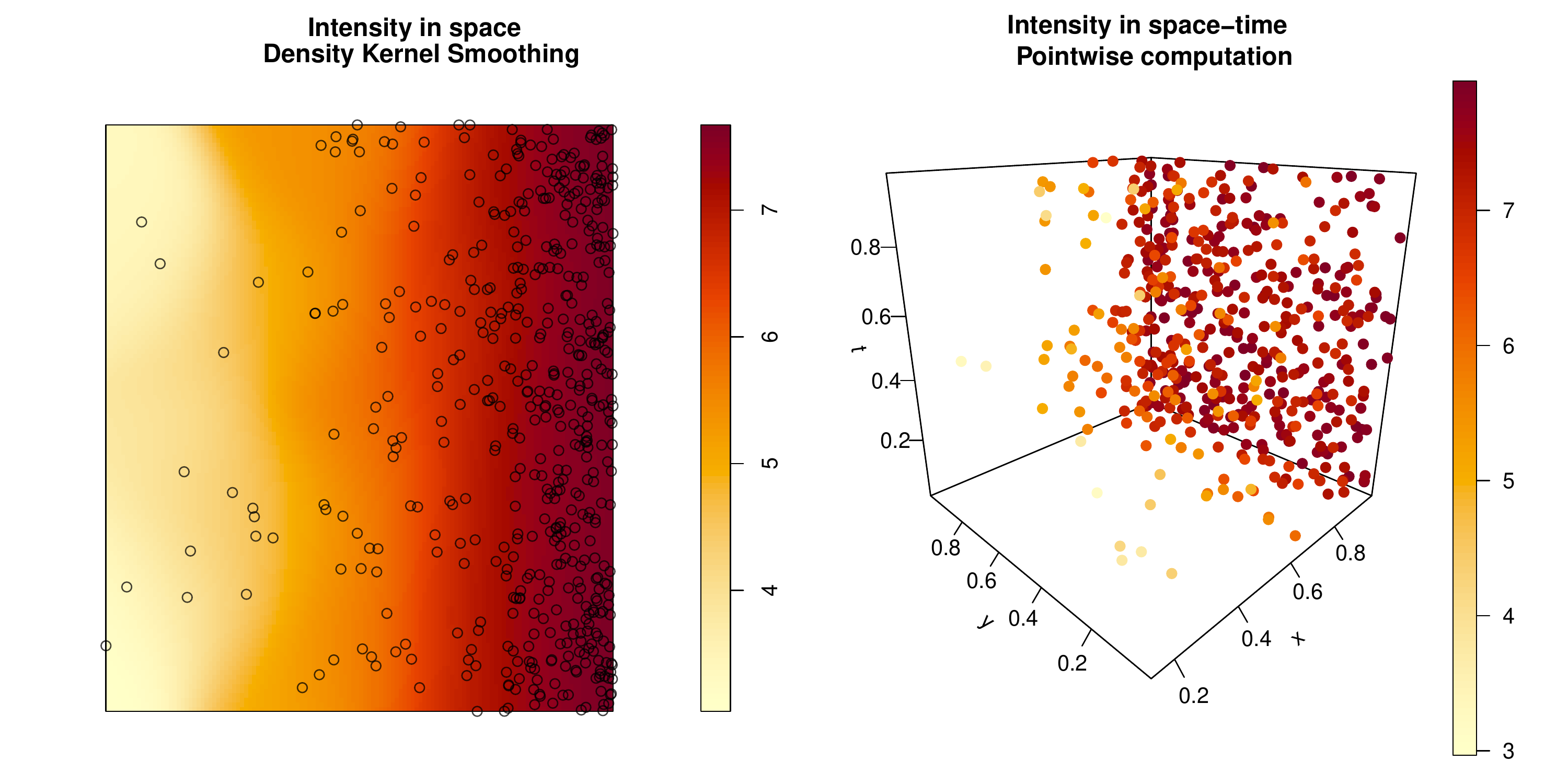}
	\caption{Output of the model fitting.}
	\label{fig:p6}
\end{figure}

\subsubsection{Local inhomogeneous spatio-temporal Poisson processes  trough local log-likelihood}

The \texttt{locstppm} function fits a Poisson process model to an observed spatio-temporal
point pattern stored in a \texttt{stp} object, that is, a Poisson model with
a set of parameters $\theta_i$ for each point $i$.
We assume that the template model is a Poisson process, with a parametric
intensity or rate function $\lambda(\textbf{u}, t; \theta_i)$  with space
and time localtions $\textbf{u} \in W,  t \in T$ and parameters $\theta_i \in \Theta.$
Estimation is performed through the fitting of a \texttt{glm} using a localised version of the quadrature scheme by \cite{berman1992approximating}, firstly introduced
in the purely spatial context by \citep{baddeley:2017local}, and in the spatio-temporal
framework by \cite{d2022locally}.

The local log-likelihood associated with the  spatio-temporal location 
 $(\textbf{v},s)$ is given by 
 $$\log L((\textbf{v},s);\theta) = \sum_i w_{\sigma_s}(\textbf{u}_i - \textbf{v}) w_{\sigma_t}(t_i - s)
 \lambda(\textbf{u}_i, t_i; \theta)  - \int_W \int_T
 \lambda(\textbf{u}, t; \theta) w_{\sigma_s}(\textbf{u}_i - \textbf{v}) w_{\sigma_t}(t_i - s) \text{d}t \text{d}u$$
 where $w_{\sigma_s}$ and $w_{\sigma_t}$ are weight functions, and 
 $\sigma_s, \sigma_t > 0$ are the smoothing bandwidths. It is not
 necessary to assume that $w_{\sigma_s}$ and $w_{\sigma_t}$
  are probability densities. For simplicity, we shall consider only kernels of fixed
 bandwidth, even though spatially adaptive kernels could also be used.
 Note that if the template model is the homogeneous Poisson process with intensity 
 $\lambda$, then the local
 likelihood estimate $\hat{\lambda}(\textbf{v}, s)$ 
 reduces to the kernel estimator of the point process intensity with
 kernel proportional to $w_{\sigma_s}w_{\sigma_t}$.
 We now use an  approximation similar to 
 $\log L(\theta) \approx
 \sum_j
 a_j
 (y_j \log \lambda(\textbf{u}_j, t_j; \theta) - \lambda(\textbf{u}_j, t_j; \theta))
 +
   \sum_j
 a_j,$
 but for the local log-likelihood associated
 with each desired location $(\textbf{v},s) \in W \times T$, that is:
 $$\log L((\textbf{v},s); \theta) \approx
 \sum_j
 w_j(\textbf{v},s)a_j
 (y_j \log \lambda(\textbf{u}_j,t_j; \theta) - \lambda(\textbf{u}_j,t_j; \theta))
 +
   \sum_j
 w_j(\textbf{v},s)a_j   ,$$
 where $w_j(\textbf{v},s) = w_{\sigma_s}(\textbf{v} - \textbf{u}_j) 
 w_{\sigma_t}(s - t_j)$.
 Basically, for each
 desired location $(\textbf{v},s)$,
  we replace the vector of quadrature weights $a_j$ by 
  $a_j(\textbf{v},s)= w_j(\textbf{v},s)a_j$ where
 $w_j (\textbf{v},s) = w_{\sigma_s}(\textbf{v} - \textbf{u}_j)w_{\sigma_t}(s - t_j)$,
  and use the GLM software to fit the Poisson regression.
 The local likelihood is defined at any location $(\textbf{v},s)$ in continuous space. 
 In practice, it is sufficient to
 consider a grid of points $(\textbf{v},s)$.
 We refer to \cite{d2022locally} for further discussion on bandwidth selection
 and on computational costs.

\begin{example}
> inh00_local <- locstppm(pin, formula = ~ 1)

> inh00_local

Homogeneous Poisson process 
with median Intensity: 7.564067

Summary of estimated coefficients 
       V1       
 Min.   :3.981  
 1st Qu.:7.291  
 Median :7.564  
 Mean   :7.316  
 3rd Qu.:7.669  
 Max.   :7.854
 
> inh01_local <- locstppm(pin, formula = ~ x)

> inh01_local

Inhomogeneous Poisson process 
with Trend: ~x

Summary of estimated coefficients 
       V1              V2        
 Min.   :1.282   Min.   :0.7667  
 1st Qu.:2.634   1st Qu.:4.5470  
 Median :3.059   Median :5.0662  
 Mean   :3.082   Mean   :5.0373  
 3rd Qu.:3.528   3rd Qu.:5.5636  
 Max.   :4.709   Max.   :6.9729 
\end{example}
\subsubsection{Log-Gaussian Cox processes estimation trough (locally weighted) joint minimum contrast}


 	In the Euclidean context, LGCPs are one of the most prominent clustering models. By specifying the intensity of the process and the moments of the underlying GRF, it is possible to estimate both the first and second-order characteristics of the process. 
	Following the inhomogeneous specification in \cite{diggle:moraga:13}, a  LGCP for a generic point in space and time has the intensity
	\begin{equation*}
		\Lambda(\textbf{u},t)=\lambda(\textbf{u},t)\exp(S(\textbf{u},t))
	\end{equation*}
where $S$ is a Gaussian process with $\mathbb{E}(S(\textbf{u},t))=\mu=-0.5\sigma^2$ and so  $\mathbb{E}(\exp{S(\textbf{u},t)})=1$ and with variance and covariance matrix $\mathbb{C}(S(\textbf{u}_i,t_i),S(\textbf{u}_j,t_j))=\sigma^2 \gamma(r,h)$ under the stationary assumption, with $\gamma(\cdot)$ the correlation function of the GRF, and $r$ and $h$ some spatial and temporal distances. Following \cite{moller1998log}, the first-order product density and the pair correlation function of an LGCP are $\mathbb{E}(\Lambda(\textbf{u},t))=\lambda(\textbf{u},t)$ and $g(r,h)=\exp(\sigma^2\gamma(r,h))$, respectively.  

The \texttt{stlgcppm} function estimates a local log-Gaussian Cox process (LGCP), following the locally weighted minimum contrast procedure introduced in \cite{d2022locally}.
Three covariances are available: separable exponential, Gneiting, and DeIaco-Cesare.
If both the first and second arguments are set to global, a log-Gaussian Cox process is fitted by means of the joint minimum contrast procedure proposed in \cite{siino2018joint}.

We may consider a separable structure for the covariance function of the GRF \citep{brix2001spatiotemporal} that has exponential form for both the spatial and the temporal components,
	\begin{equation}
		\mathbb{C}(r,h)=\sigma^2\exp \bigg(\frac{-r}{\alpha}\bigg)\exp\bigg(\frac{-h}{\beta}\bigg),
		\label{eq:cov}
	\end{equation}
where $\sigma^2$ is the variance, $\alpha$ is the scale parameter for the spatial distance and $\beta$ is the scale parameter for the temporal one.
The exponential form is widely used in this context and nicely reflects the decaying correlation structure with distance or time.\\
Moreover, we may consider a non-separable covariance of the GRF useful to describe
more general situations.
Following the parametrisation in \cite{schlather2015analysis}, Gneiting covariance function \citep{gneiting2006geostatistical} can be written as
$$
		\mathbb{C}(r,h) = (\psi(h) + 1)^{ - d/2} \varphi \bigg( \frac{r}{\sqrt{\psi(h) + 1}}  \bigg) \qquad r \geq 0,  \quad h \geq 0, 
$$
where $\varphi(\cdot)$ is a complete monotone function associated to
the spatial structure, and $\psi(\cdot)$ is a positive function with a
completely monotone derivative associated to the temporal
structure of the data. For example, the choice $d = 2$,
$\varphi(r)=\sigma^2 \exp ( - (\frac{r}{\alpha})^{\gamma_s})$ and 
$\psi(h)=((\frac{h}{\beta})^{\gamma_t} + 1)^{\delta/\gamma_t}$
yields to the parametric family
	\begin{equation}
		\mathbb{C}(r,h) = \frac{\sigma^2}{((\frac{h}{\beta})^{\gamma_t} + 1)^{\delta/\gamma_t}} \exp \Biggl( - \frac{(\frac{r}{\alpha})^{\gamma_s}}{((\frac{h}{\beta})^{\gamma_t} + 1)^{\delta/(2\gamma_t)}} \Biggl),
		\label{eq:nonsep}
	\end{equation}
where $\alpha > 0$ and $\beta > 0$ are scale parameters of space and time, $\delta$ takes values in $(0, 2]$, and $\sigma^2$ is the variance.\\
Another parametric covariance implemented belongs to the Iaco-Cesare family \citep{de2002fortran,de2002nonseparable}, and there is a wealth of covariance families that could well be used for our purposes.

 Following \cite{siino2018joint}, the second-order
 parameters $\boldsymbol{\psi}$ are found by minimising 
 $$M_J\{ \boldsymbol{\psi}\}=\int_{h_0}^{h_{max}} \int_{r_0}^{r_{max}} \phi(r,h) \{\nu[\hat{J}(r,h)]-\nu[J(r,h;\boldsymbol{\psi})]\}^2 \text{d}r \text{d}h,$$
 where $\phi(r, h)$ is a weight that depends on the space-time
 distance and $\nu$ is a transformation function. 
 They suggest $\phi(r,h)=1$ and $\nu$ as
 the identity function, while $r_{max}$ and $h_{max}$ are selected as 1/4
 of the maximum observable spatial and temporal distances.\\
 Following \cite{d2022locally}, we can fit a localised version of the LGCP, 
 that is,  obtain a
 vector of parameters $\boldsymbol{\psi}_i$ for each point $i$, by
 minimising 
   $$M_{J,i}\{ \boldsymbol{\psi}_i \}=\int_{h_0}^{h_{max}}\int_{r_0}^{r_{max}}
    \phi(r,h) \{ \nu[\bar{J}_i(r,h)]-\nu[J(r,h;\boldsymbol{\psi})]\}^2 \text{d}r \text{d}h
 \qquad \text{with} \qquad 
    \bar{J}_i(r,h)= \frac{\sum_{i=1}^{n}\hat{J}_i(r,h)w_i}{\sum_{i=1}^{n}w_i}$$
    is the average of the local functions
 $\hat{J}_i(r,h)$, weighted by some point-wise kernel estimates. 
   In particular, we consider $\hat{J}_i(\cdot)$ as the local
 spatio-temporal pair correlation function \citep{gabriel:rowlingson:diggle:2013} documented in \texttt{LISTAhat}.

The \texttt{print} and \texttt{summary} functions give the main information on the fitted model. In case of local parameters (both first- and second-order), the summary function contains information on their distributions.
Next, we perform and example with a complex seismic point pattern.
\begin{example}
> data("greececatalog")
\end{example}
If both first and second arguments are set to "global", a log-Gaussian Cox process is fitted by means of the joint minimum contrast.
\begin{example}
> lgcp1 <- stlgcppm(greececatalog, formula = ~ 1, first = "global", second = "global")
> lgcp1

Joint minimum contrast fit 
for a log-Gaussian Cox process with 
global first-order intensity and 
global second-order intensity 
--------------------------------------------------
Homogeneous Poisson process 
with Intensity: 0.00643

Estimated coefficients of the first-order intensity: 
(Intercept) 
     -5.046 
--------------------------------------------------
Covariance function: separable 

Estimated coefficients of the second-order intensity: 
  sigma   alpha    beta 
  6.989   0.225 156.353 
--------------------------------------------------
Model fitted in 1.014 minutes
 \end{example}
If first = "local", local parameters for the first-order intensity are provided. In this case, the summary function contains information on their distributions. 
\begin{example}
> lgcp2 <- stlgcppm(greececatalog, formula = ~ x, first = "local", second = "global")
> lgcp2

Joint minimum contrast fit 
for a log-Gaussian Cox process with 
local first-order intensity and 
global second-order intensity 
--------------------------------------------------
Inhomogeneous Poisson process 
with Trend: ~x

Summary of estimated coefficients of the first-order intensity 
  (Intercept)           x           
 Min.   :-6.400   Min.   :-0.90689  
 1st Qu.:-2.526   1st Qu.:-0.38710  
 Median : 2.333   Median :-0.26876  
 Mean   : 2.153   Mean   :-0.26744  
 3rd Qu.: 5.070   3rd Qu.:-0.06707  
 Max.   :16.323   Max.   : 0.10822  
--------------------------------------------------
Covariance function: separable 

Estimated coefficients of the second-order intensity: 
 sigma  alpha   beta 
 2.612  0.001 36.415 
--------------------------------------------------
Model fitted in 3.634 minutes
\end{example}
The \texttt{plot} function shows the fitted intensity, displayed 
 both in space (by means of a density kernel smoothing) and in space and time. In the case of local covariance parameters, the function returns the mean of the random intensity, displayed both in space (by means of a density kernel smoothing) and in space and time.
The \texttt{localsummary.stlgcppm} function breaks up the contribution of the local estimates to the fitted intensity, by plotting the overall intensity and the density kernel smoothing of some artificial intensities obtained by imputing the quartiles of the local parameters' distributions.
Finally, the function \texttt{localplot.stlgcppm} function plots the local estimates. In the case of local covariance parameters, the function displays the local estimates of the chosen covariance function.
\begin{figure}[H]
	\centering
	\includegraphics[width=\textwidth]{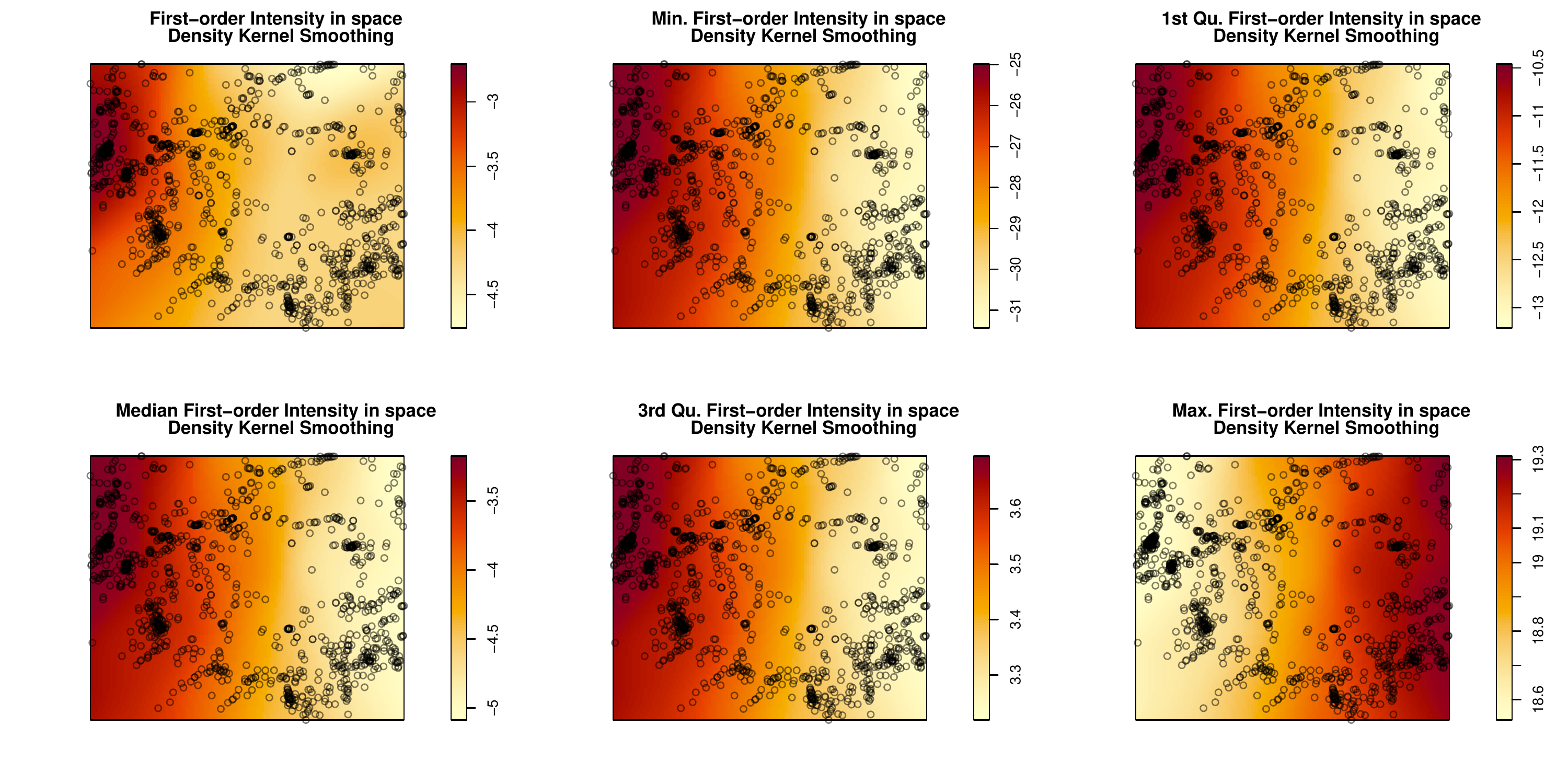}
	\caption{Output of the \texttt{localsummary} function.}
	\label{fig:p7}
\end{figure}
\begin{figure}[H]
	\centering
 \includegraphics[width=\textwidth]{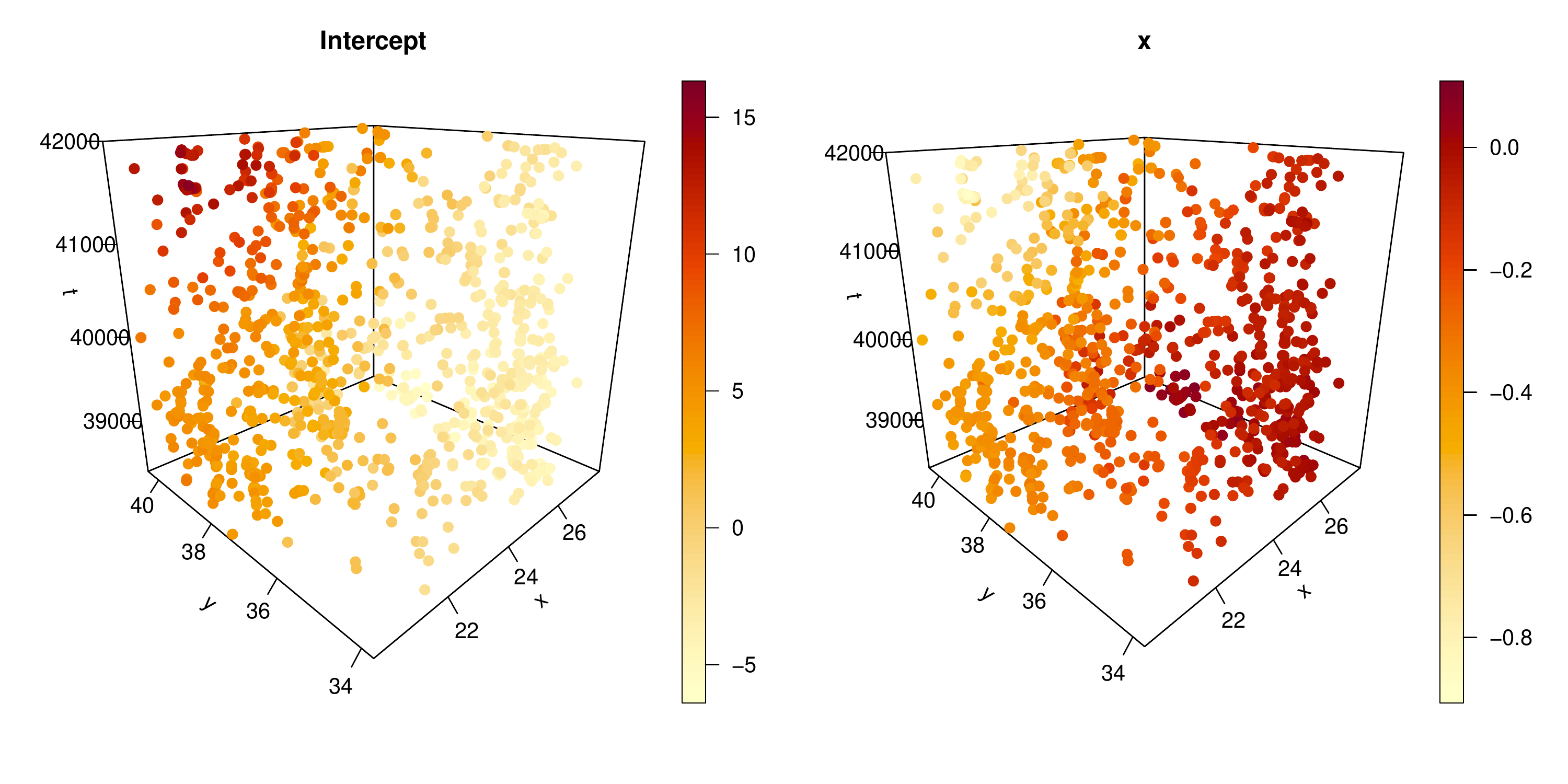}
	\caption{Estimated local coefficients.}
	\label{fig:p75}
\end{figure}







\section{Diagnostics}\label{sec:diag}

Inhomogeneous second-order statistics can be constructed and used for assessing the goodness-of-fit of fitted first-order intensities.
Nevertheless, it is a widespread practice in the statistical analysis of spatial and spatio-temporal point pattern data  primarily comparing the data with a homogeneous Poisson process, which is generally the null model in applications for the fitted model. Indeed, when dealing with diagnostics in point processes, often two steps are needed: the transformation of data into residuals (thinning or rescaling \citep{schoenberg2003multidimensional}) and the use of tests to assess the consistency of the residuals with the homogeneous Poisson process \citep{adelfio:schoenberg:09}. Usually, second-order statistics estimated for the residual process (i.e. the result of a thinning or rescaling procedure) are analysed.
Essentially, to each observed point a weight inversely proportional to the conditional intensity at that point is given. This method was adopted by \cite{veen2006assessing} in constructing a weighted version of the $K$-function of \cite{ripley1977markov}; the resulting weighted statistic is in many cases more powerful than residual methods \citep{veen2006assessing}. \\
The spatio-temporal inhomogeneous version of the $K$-function in \eqref{eq:k} is given by \cite{gabriel2009second} as
\begin{equation}
        \hat{K}_{I}(r,h)=\frac{  \vert W  \vert  \vert T  \vert }{n(n-1)}\sum_{i=1}^n \sum_{j > i} \frac{I( \vert  \vert \textbf{u}_i-\textbf{u}_j  \vert  \vert \leq r,\vert t_i-t_j\vert  \leq h)}{\hat{\lambda}(\textbf{u}_i,t_i)\hat{\lambda}(\textbf{u}_j,t_j)},
    \label{eq:kinh}
\end{equation}
where  $\lambda(\cdot,\cdot)$ is the first-order intensity at an arbitrary point.
We know that $\mathbb{E}[\hat{K}_{I}(r,h)]=\pi r^2 h$, that is the same as the expectation of $\hat{K}(r,h)$ in \eqref{eq:k}, when the intensity used for the weighting is the true generator model.
This is a crucial result that allows the use of the weighted estimator $\hat{K}_{I}(r,h)$ as a diagnostic tool, for assessing the goodness-of-fit of spatio-temporal point processes with generic first-order intensity functions.
 Indeed, if the weighting intensity function  is close to the true one $\lambda(\textbf{u},t)$, the expectation of $\hat{K}_{I}(r,h)$ should be close to $\mathbb{E}[\hat{K}(r,h)]=\pi r^2 h$ for the Poisson process. For instance, values $\hat{K}_{I}(r,h)$ greater than $\pi r^{2} h$ indicates that the fitted model is not appropriate, since the distances computed among points exceed the Poisson theoretical ones.

The \texttt{globaldiag} function performs global diagnostics of a model fitted for the first-order intensity of an spatio-temporal point pattern, using the spatio-temporal inhomogeneous K-function \citep{gabriel2009second} documented by the function \texttt{STIKhat} of the \textbf{stpp} package \citep{stpp}.
It can also perform global diagnostics of a model fitted for the first-order intensity of an spatio-temporal point pattern on a linear network, by means of the spatio-temporal inhomogeneous K-function on a linear network \citep{moradi2020first} documented by the function \texttt{STLKinhom} of the \textbf{stlnpp} package \citep{stlnpp}.
They both return the plots of the inhomogeneous K-function weighted by the provided intensity to diagnose, its theoretical value, and their difference.

\begin{example}
> globaldiag(greececatalog, lgcp1$l)
[1] "Sum of squared differences =  318213525081.852"

> globaldiag(greececatalog, lgcp2$l)
[1] "Sum of squared differences =  147029066885.741"
\end{example}
\begin{figure}[H]
	\centering
	\includegraphics[width=\textwidth]{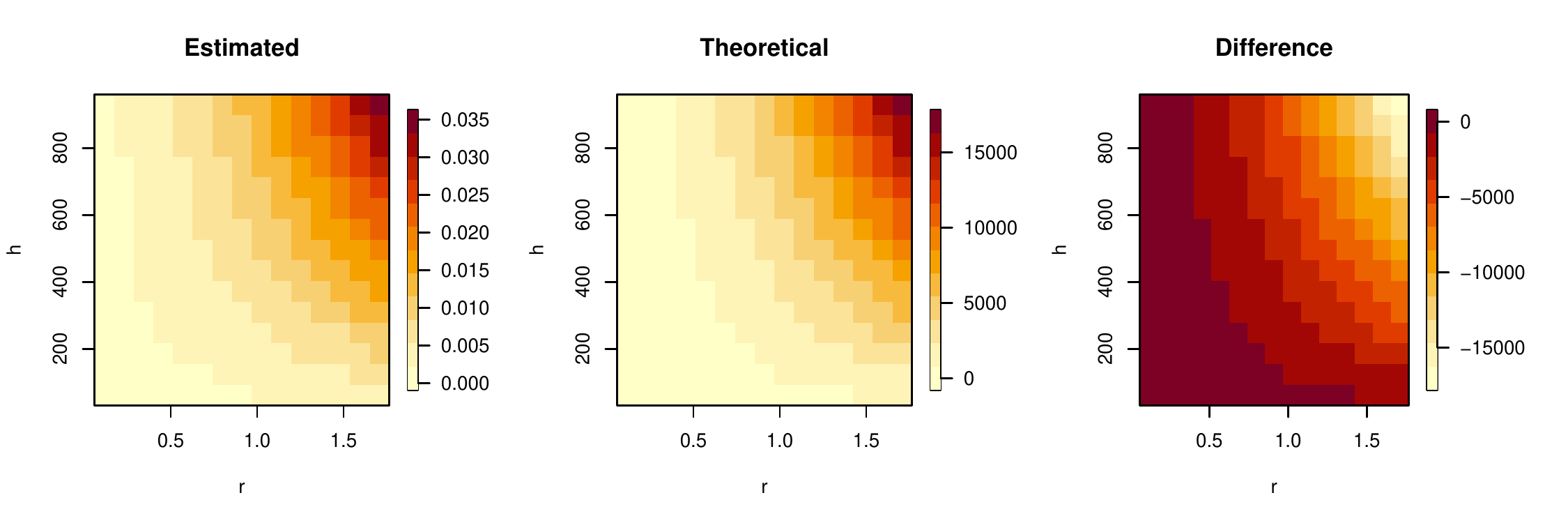}\\
 \vspace{-.5cm}
 \includegraphics[width=\textwidth]{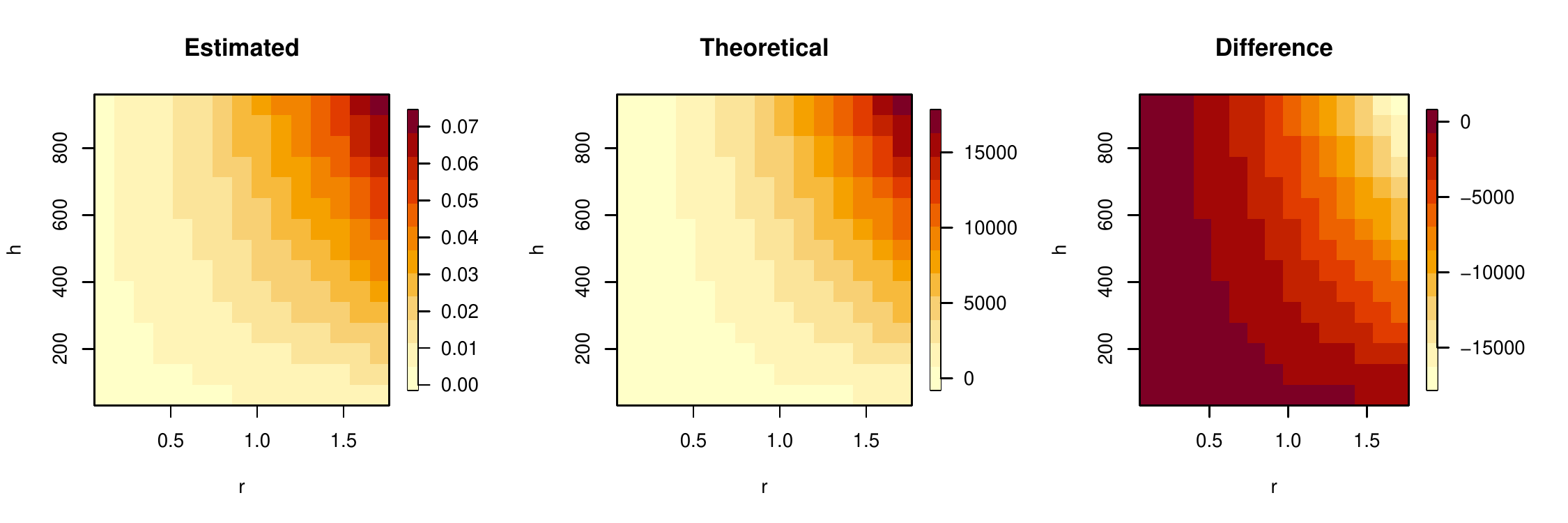}
	\caption{Output of the global diagnostics for the two fitted LGPCs.}
	\label{fig:p9}
\end{figure}

Moving to the local diagnostics, \cite{adelfio2020some} derived the expectation of the local
 inhomogeneous spatio-temporal K-function, under the Poisson case: 
 $\mathbb{E}[\hat{K}^i(r,h) ]= \pi r^ 2 h.$
 Moreover, they found that when the local estimator is weighted by the true
 intensity function,
  its
 expectation, $\mathbb{E}[\hat{K}_{I}^i(r,h)]$, is the same as the expectation of
 $\hat{K}^i(r,h)$.
 These results motivate the usage of such local estimator
 $\hat{K}_{I}^i(r,h)$ as a diagnostic tool for general spatio-temporal
 point processes for assessing the goodness-of-fit of spatio-temporal
 point processes of any generic first-order
 intensity function $\lambda$.
 Indeed, if the estimated intensity function
 used for weighting in our proposed LISTA functions is
 the true one, then the LISTA functions should behave as the
 corresponding ones of a homogeneous Poisson process,
 resulting in small discrepancies between the two.
 Therefore, this function computes such discrepancies
 by means of the $\chi_i^2$ values, obtained following the expression
 $$ \chi_i^2=\int_L \int_T \Bigg( 
     \frac{\big(\hat{K}^i_{I}(r,h)- \mathbb{E}[\hat{K}^i(r,h) ] 
     \big)^2}{\mathbb{E}[\hat{K}^i(r,h) ]} 
     \Bigg) \text{d}h \text{d}r ,$$
     one for each point in the point pattern. 
 Basically, departures of the LISTA functions $\hat{K}_{I}^i(r,h)$ from
 the Poisson expected value $rh$ directly suggest the unsuitability of
 the intensity function $\lambda(\cdot)$ used in the weighting of the
 LISTA functions for that specific point. This can be referred to as  an \textit{outlying point}.
 Given that \cite{dangelo2021local} proved the same results for the network case, 
 that is,
 $\mathbb{E}[\hat{K}_{L}^i(r,h) ]= rh$ and
  $\mathbb{E}[\hat{K}_{L,I}^i(r,h) ]=\mathbb{E}[\hat{K}_{L}^i(r,h) ]$
 when $\hat{K}_{L,I}^i(r,h)$ is weighted by the true intensity function,
 we implemented the same above-mentioned diagnostics procedure to work on
 intensity functions fitted on spatio-temporal point patterns occurring on 
 linear networks.
 Note that the Euclidean procedure is implemented by means of the 
 local K-functions of 
    \cite{adelfio2020some}, documented in
  \texttt{KLISTAhat} of the \textbf{stpp} package \citep{gabriel:rowlingson:diggle:2013}.
 The network case uses  the local K-functions on networks \citep{dangelo2021local},
  documented
 in  \texttt{localSTLKinhom}.
 
The \texttt{localdiag} function performs local diagnostics of a model fitted for the first-order intensity of an spatio-temporal point pattern, by means of the local spatio-temporal inhomogeneous K-function \citep{adelfio2020some} documented by the function KLISTA of the \textbf{stpp} package \citep{gabriel:rowlingson:diggle:2013}.
It returns the points identified as outlying following the diagnostics procedure on individual points of an observed point pattern, as introduced in \cite{adelfio2020some}.
The points resulting from the local diagnostic procedure provided by this function can be inspected via the \texttt{plot}, \texttt{print}, \texttt{summary}, and \texttt{infl} functions.
\texttt{localdiag} is also able to perform local diagnostics of a model fitted for the first-order intensity of an spatio-temporal point pattern on a linear network, by means of the local spatio-temporal inhomogeneous K-function on linear networks \cite{dangelo2021assessing} documented by the function \texttt{localSTLKinhom}.
It returns the points identified as outlying following the diagnostics procedure on individual points of an observed point pattern, as introduced in \cite{adelfio2020some}, and applied in \cite{dangelo2021local} for the linear network case.

\begin{example}
> set.seed(12345)
> stlp1 <- rETASlp(cat = NULL, params = c(0.078915 / 2, 0.003696,  0.013362,  1.2,
                                         0.424466,  1.164793),
                  betacov = 0.5, m0 = 2.5, b = 1.0789, tmin = 0, t.lag = 200,
                  xmin = 600, xmax = 2200, ymin = 4000, ymax = 5300,
                  iprint = TRUE, covdiag = FALSE, covsim = FALSE, L = chicagonet)

> res <- localdiag(stlp1, intensity = density(as.stlpp(stlp1),  at = "points"))
> res  

Points outlying from the 0.95 percentile
of the anaysed spatio-temporal point pattern on a linear network 
--------------------------------------------------
Analysed pattern X: 65 points 
4 outlying points

> plot(res)
> infl(res)
\end{example}
\begin{figure}[H]
	\centering
	\includegraphics[width=\textwidth]{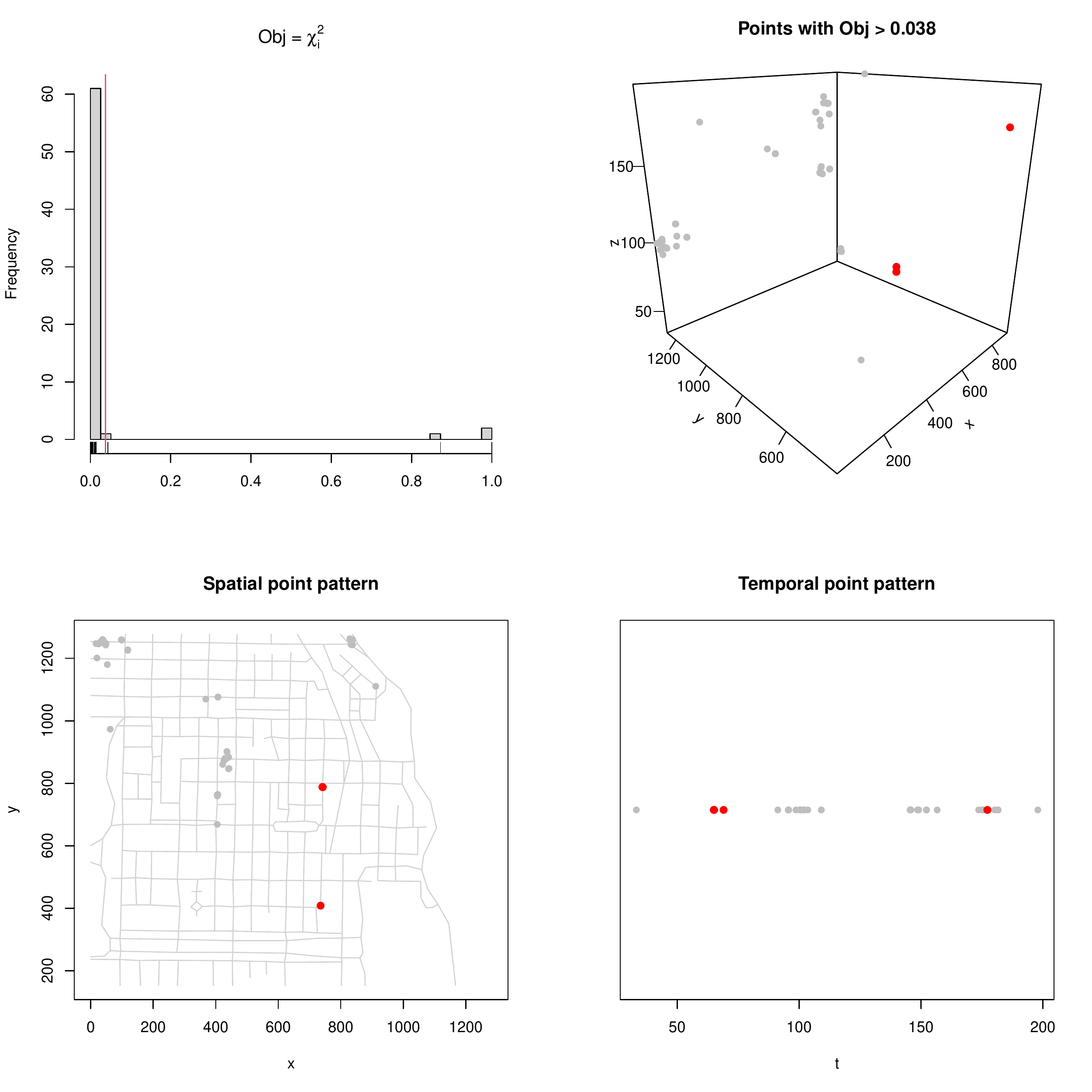}
	\caption{Output of the local diagnostics via the \texttt{plot.localdiag} function.}
	\label{fig:p12}
\end{figure}
\begin{figure}[H]
	\centering
 	\includegraphics[width=.95\textwidth]{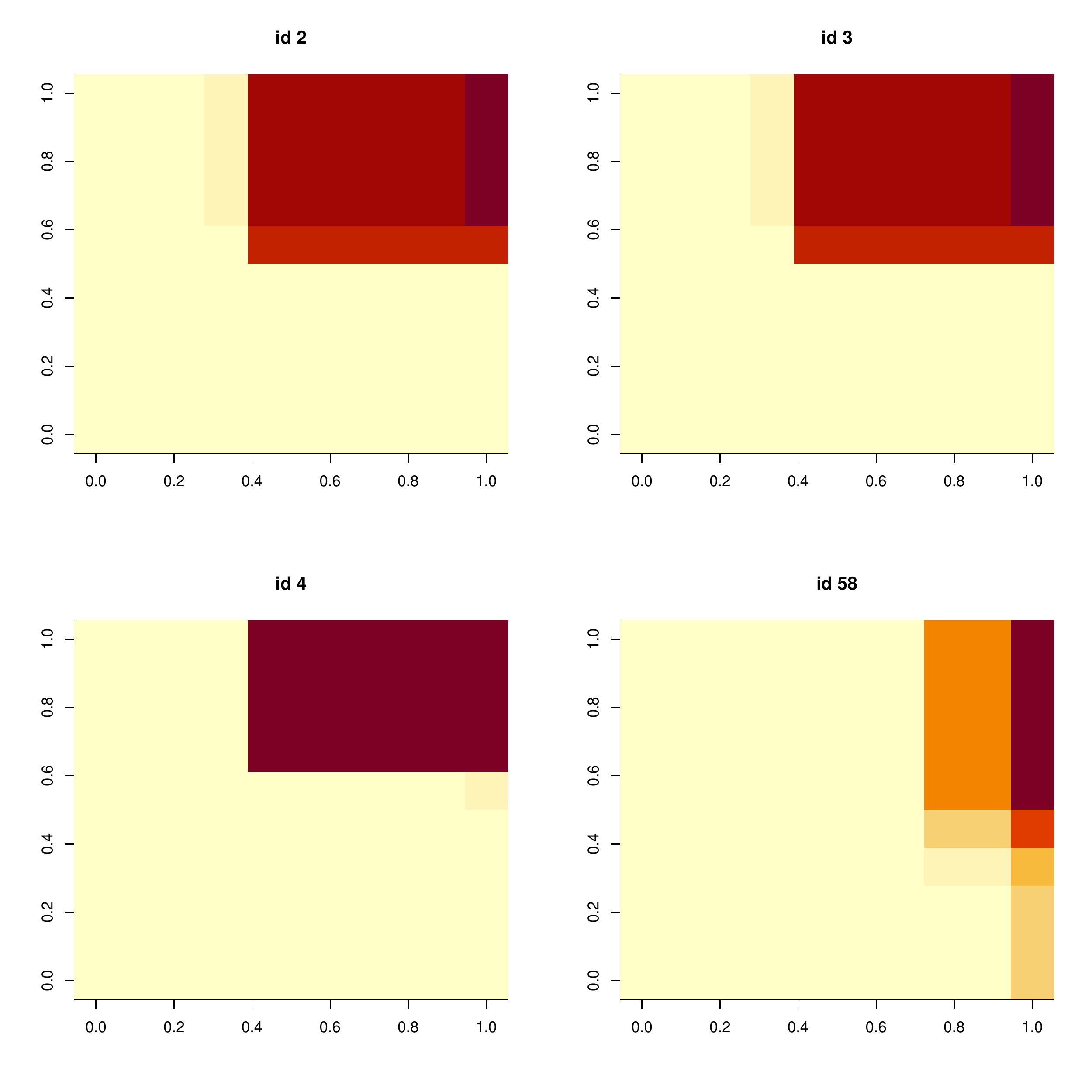}
	\caption{Output of the local diagnostics via the \texttt{infl.localdiag} function.}
	\label{fig:p13}
\end{figure}

\section{Conclusions}\label{sec:concl}

This work has introduced the   \textbf{stopp} \texttt{R} package, which deals with spatio-temporal point processes occurring either on the Euclidean space or on some specific linear networks, such as streets of a city.

The package includes functions for summarizing, plotting, and performing various analyses on point processes; these functions mostly use the approaches suggested in a few recent works in scientific literature. Modelling, statistical inference, and simulation difficulties on spatio-temporal point processes on Euclidean space and linear networks, with a focus on their local properties, are the core topics of such research and the package in turn.

To start with, we set the notation for spatio-temporal point processes that can occur in both linear networks and Euclidean spaces. After that, we went over the main methods implemented in the  \textbf{stopp}  package for dealing with simulations, data, and objects in point processes. After having recalled the definition of Local Indicators of Spatio-Temporal Association (LISTA) functions, we have moved to introduce the new functions that compute the LISTAs on linear networks.  We then illustrated functions to run a local test to evaluate the local differences between two point patterns occurring on the same metric space.  Moreover, many examples of the models included in the package are provided. These examples include: models for separable Poisson processes on both Euclidean space and networks, global and local non-separable inhomogeneous Poisson processes, and LGCPs. Then, techniques for performing both global and local diagnostics on such models (but not limited to those only)  for point patterns on linear networks and planar  spaces are provided. 

The package tools are not exhaustive. This work represents the creation of a toolbox for different kinds of spatio-temporal analyses to be performed on observed point patterns, following the growing stream of literature on point process theory.
The presented work contributes to the existing literature by framing many of the most widespread methods for the analysis of spatio-temporal point processes into a unique package, which is intended to foster many further extensions.

\bibliography{BBB}

\end{document}